\documentclass{JHEP3}
\pdfoutput=1
\usepackage{amsmath}
\usepackage{graphics}

\usepackage{epsfig}
\usepackage{amssymb}
\usepackage[active]{srcltx}

\def\({\left(}
\def\){\right)}
\def\[{\left[}
\def\]{\right]}
\def\<{\langle}
\def\>{\rangle}
\def\b{\bullet}
\def\lv{\left|}
\def\rv{\right|}

\def\und{\underline}
\def\ie{{\it i.e.,} }
\def\m{\overline{m}}
\def\ap{\alpha}
\def\bt{\beta}
\def\gm{\gamma}
\def\de{\delta}
\def\la{\lambda}
\def\tla{\tilde{\lambda}}

\newcommand{\be}{\begin{equation}}
\newcommand{\ee}{\end{equation}}
\newcommand{\bal}{\begin{aligned}}
\newcommand{\eal}{\end{aligned}}
\newcommand{\barr}{\begin{array}}
\newcommand{\earr}{\end{array}}

\newcommand{\labell}[1]{\label{#1}}

\title{On Multi-step BCFW Recursion Relations}

\author{Bo Feng$^{ab}$, Junjie Rao$^{a}$, Kang Zhou$^{a}$\footnote{Emails:
b.feng@cms.zju.edu.cn, raojunjie@zju.edu.cn, 11236072@zju.edu.cn} \\
{$^a$\small Zhejiang Institute of Modern Physics, Zhejiang University, Hangzhou, 310027, P. R. China \\
$^b$\small Center of Mathematical Science, Zhejiang University, Hangzhou, 310027, P. R. China \\}}

\abstract{In this paper, we extensively investigate the new algorithm known as the multi-step BCFW
recursion relations. Many interesting mathematical properties are found and understanding these aspects,
one can find a systematic way to complete the calculation of amplitude after finite, definite
steps and get the correct answer, without recourse to any specific knowledge from field theories, besides
mass dimension and helicities. This process consists of the pole concentration and
inconsistency elimination. Terms that survive inconsistency elimination cannot be determined by
the new algorithm. They include polynomials and their generalizations,
which turn out to be useful objects to be explored.
Afterwards, we apply it to the Standard Model plus gravity to illustrate its power and limitation.
Ensuring its workability, we also tentatively discuss how to improve its efficiency by reducing the steps.}


\begin{document}
\maketitle

\newpage
\section{Introduction}
\labell{intro}

In the past decade, the BCFW recursion relation \cite{Britto:2004ap, Britto:2005fq}
had been an efficient on-shell method to calculate tree-level
scattering amplitudes. Pedagogical reviews on this topic can be found in \cite{Feng:2011np, Elvang:2013cua}.
Still, it encounters certain difficulties when there exists
no `good' deformation as those found in \cite{ArkaniHamed:2008yf, Cheung:2008dn},
\ie the real amplitude does not vanish under the large $z$ limit, where $z$ is the deformation parameter.
The recursion relation then fails to capture a residual part called the boundary term,
which corresponds to the residue at infinity of the deformed amplitude.

Many related studies have been achieved including:
introducing auxiliary fields to eliminate boundary terms \cite{Benincasa:2007xk, Boels:2010mj},
analyzing Feynman diagrams to isolate boundary terms \cite{Feng:2009ei, Feng:2010ku, Feng:2011twa},
expressing boundary terms as roots of amplitudes \cite{Benincasa:2011kn, Benincasa:2011pg, Feng:2011jxa},
collecting factorization limits to interpolate boundary terms \cite{Zhou:2014yaa} and
using other deformations for better large $z$ behavior \cite{Cheung:2015cba}.

Recently, a new algorithm named as the
multi-step BCFW recursion relations \cite{Feng:2014pia}
was established to tackle this problem universally. Its major idea of using auxiliary deformations
can be traced back to \cite{Berger:2006ci}, while the latter aims for one-loop amplitudes.
This approach considerably widens the category of quantum field theories of
solvable tree amplitudes by using BCFW deformations only \cite{Jin:2014qya}.
However, some common puzzles encountered in practice still lacks a formal study. One core question is:
How to reach the correct answer within finite, definite steps, if an amplitude is solvable by the algorithm?

In this paper, we will first explore multi-step BCFW recursion relations
by investigating the algebra of BCFW deformation generators and the commutativity of constant extractions.
Next, we will seek for a universal approach to reach the answer and ensure that it is correct.
This safety promise relies on very little knowledge of a particular QFT, besides mass dimension and helicities,
hence the algorithm is expected to be able to solve for all massless tree amplitudes,
with certain limitation as addressed below.

It is well known that on-shell methods heavily rely on factorization properties of amplitudes,
and the latter is a reflection of locality and unitarity. These properties are mathematically implemented
on poles of amplitudes and their residues. For amplitudes that admit polynomials, no on-shell methods so far
can fix this ambiguity. One can list all possible forms of polynomials as basis, but to determine the
coefficients will unfavorably call for more traditional means such as Feynman rules.
In this work, we will clarify the applicable range of multi-step BCFW recursion relations and explore
all possible forms of polynomials and their generalized cousins called pseudo polynomials
and saturated fractions. The latter two objects can be fixed by other types of deformations,
and having them fully identified is in fact useful.

The paper is organized as follows.
In section \ref{sec2}, we review the multi-step BCFW recursion relations and explore the commutativity
of constant extractions. In section \ref{sec3}, we propose the systematic process to calculate
amplitudes after finite, definite steps, and clarify its applicable range and limitation.
In section \ref{sec4}, we apply it to the (massless) Standard Model plus gravity
to demonstrate its workability.

\section{Multi-step BCFW Recursion Relations}
\labell{sec2}

In this section, we briefly review the multi-step BCFW recursion relations, in the novel language of
extraction operators. After that, the commutativity of constant extractions will be explored.

\subsection{Extraction operators}

For a general BCFW deformation $\<a_i|b_i]$ (only two legs are shifted for each $i$), namely
\be
\la_{a_i}\to\la_{a_i}-z_i\la_{b_i},~\tla_{b_i}\to\tla_{b_i}+z_i\tla_{a_i},
\ee
let's define two operations on an amplitude-like rational function $R(\la_i,\tla_i)$
via\footnote{$P_i$ and $C_i$ used here are identical to $\mathcal{P}^{\und{i}}$ and
$\mathcal{C}^{\und{i}}$ in the appendix of \cite{Feng:2014pia}.}
\be
P_i[R]\equiv-\sum_{\textrm{finite}}\oint\frac{dz_i}{z_i}
R(\la_{a_i}-z_i\la_{b_i},\tla_{b_i}+z_i\tla_{a_i}),~
C_i[R]\equiv\oint_\infty\frac{dz_i}{z_i}R(\la_{a_i}-z_i\la_{b_i},\tla_{b_i}+z_i\tla_{a_i}),
\ee
where $P_i$ and $C_i$ are the pole and constant `extraction operators', which capture
residues at finite locations \textit{except} zero and infinity respectively.
For a real amplitude $A$, $P_i$ can capture its physical poles only.
But a general $R$, such as $P_iA\equiv P_i[A]$ or $C_iA\equiv C_i[A]$, may also contain spurious poles,
which is well known. Therefore the detectable poles (those which have dependence on $z_i$,
as defined in \cite{Feng:2014pia}) at finite locations can be either physical or spurious.

By definition $P_i+C_i=I$, where $I$ is the identity operator. When we calculate an amplitude,
starting by the 0th step, the amplitude is unknown, so is the $C_0$ operation.
However, the $P_0$ operation represents exactly the BCFW recursion relation,
hence we actually reconstruct this part by employing factorization properties,
rather than manipulating the unknown amplitude. Conventionally, $C_0$ is called the boundary term
with respect to $P_0$, which will be dissected into many parts to be determined.
The dissection means, by expanding $I$ for $(n+1)$ times repeatedly, we have
\be
I=P_n+C_nP_{n-1}+\ldots+C_nC_{n-1}\cdots
C_2P_1+C_nC_{n-1}\cdots C_2C_1P_0+C_nC_{n-1}\cdots C_2C_1C_0,
\ee
note that $I$ always acts on $A$ implicitly.
If the final boundary term $C_nC_{n-1}\cdots C_2C_1C_0$ vanishes, we have
\be
I=P_n+C_nP_{n-1}+\ldots+C_nC_{n-1}\cdots C_2P_1+C_nC_{n-1}\cdots C_2C_1P_0. \labell{eq-1}
\ee
This identity formally represents the `multi-step BCFW recursion relations'.
Importantly, the workability of this multi-step approach relies on the existence of a sequence
of deformations numbered by $0,1,\ldots,n$ for which $C_nC_{n-1}\cdots C_2C_1C_0=0$.
The latter is the key condition we will mainly focus on.

The operators above have a general algebraic property, namely the projectivity:
\be
C_iC_i=C_i,
\ee
to prove this, we first explicitly expand the deformed $R$ as\footnote{In practice,
one can use the `Apart' function in Mathematica
to separate the pole and regular terms with respect to $z$.}
\be
R(z_i)=\sum_k\frac{b_{0k}+b_{1k}z_i}{\(a_{0k}+a_{1k}z_i+a_{2k}z_i^2\)^{d_k}}
+c_0+\sum_lc_lz_i^l, \labell{eq-10}
\ee
with $d_k\geq1$. In the expansion, when $a_{2k}$ vanishes, $b_{1k}$ must also vanish, otherwise
a linear recombination of the numerator can further lower $d_k$ by one\footnote{The $z^2$
term in the denominator can only originate
from spurious pole $\<i|K|j]$, where $K$ contains at least two external momenta
other than $i,j$, when it is deformed by $\<i|j]$. All other physical poles can at most contribute
terms linear in $z$ under one BCFW deformation.}.
Now observe that performing the same deformation twice is equivalent to
replacing $z_i$ by $(z_i+z'_i)$, as
\be
R(z_i,z'_i)=R(z_i+z'_i)
=\sum_k\frac{b_{0k}+b_{1k}(z_i+z'_i)}{\(a_{0k}+a_{1k}(z_i+z'_i)+a_{2k}(z_i+z'_i)^2\)^{d_k}}
+c_0+\sum_lc_l(z_i+z'_i)^l,
\ee
then
\be
\oint_\infty\frac{dz'_i}{z'_i}\oint_\infty\frac{dz_i}{z_i}R(z_i,z'_i)
=\oint_\infty\frac{dz'_i}{z'_i}\(c_0+\sum_lc_l{z'_i}^l\)=c_0,
\ee
hence $C_iC_iR=C_iR=c_0$. By using $P_i=I-C_i$ it is trivial to find that
\be
P_iP_i=P_i,~C_iP_i=P_iC_i=0.
\ee
Besides projectivity, a more intricate property is the commutativity:
\be
C_iC_j=C_jC_i,
\ee
which demands certain condition, as will be investigated shortly.
If it holds, again with $P_i=I-C_i$ one can find that
\be
P_iP_j=P_jP_i,~C_iP_j=P_jC_i. \labell{eq-2}
\ee
When all $C$'s are chosen to commute with each other in the expansion \eqref{eq-1}
for a particular amplitude, each term is `orthogonal' to the others.
This orthogonality has a nice meaning: Each term contains non-overlapping pole terms,
consequently one can capture all pole terms step by step \textit{without} checking
whether the previous parts are disturbed by new operations.
While commutativity may considerably simplify the calculation, it is obviously not necessary
for \eqref{eq-1} to work.

One last digression is when we do practical calculations, it is convenient to use $P_i+C_i=I$
to switch between $P_i$ and $C_i$, depending on which operation is easier. To check the equivalence between
two visually different expressions, in appendix \ref{app1} we introduce a simple trick to solve all constraints
and get a set of independent kinematic variables. This trick can uniquely fix
the form of an expression no matter by which means it is obtained
(it is better to use this trick with a computer algebra program).

\subsection{Deformation generator algebra}

Now we begin to explore the commutativity of $C$'s, which can be decomposed into the
commutativity at integrand level and at integral level. The former is encoded in two successive
deformations, and the latter is encoded in two successive contour integrals, which will use
Laurent expansion in $w=1/z$. Before this, we need to first
study the BCFW deformation generators and their algebra.

Let's define the BCFW deformation generator with respect to $\<i|j]$ as
\be
D_{\<i|j]}\equiv-\la_j^\ap\frac{\partial}{\partial\la_i^\ap}
+\tla_i^{\dot{\ap}}\frac{\partial}{\partial\tla_j^{\dot{\ap}}},
\ee
then the familiar BCFW deformation becomes
\be
\exp\(zD_{\<i|j]}\)R(\la_i,\tla_j)=R(\la_i-z\la_j,\tla_j+z\tla_i).
\ee
Although by default the spinorial partial derivatives treat all spinors as independent,
we must also impose the momentum conservation constraint on real amplitudes.
Without doubt, this constraint will affect the independence of spinorial partial derivatives,
but it will \textit{not} affect the commutator algebra of $D_{\<i|j]}$.
Below we will provide a simple argument.

Note that any $D_{\<i|j]}$ automatically annihilates the sum of all external momenta, \ie
\be
D_{\<i|j]}\sum p=D_{\<i|j]}(\la_i\tla_i+\la_j\tla_j)=0,
\ee
so we claim that momentum conservation is a \textit{trivial} constraint.
To get some intuition, one can consider a spherical surface,
for which any rotation generator, say $L_{xy}$, annihilates the constraint
\be
x^2+y^2+z^2=r^2.
\ee
To parameterize one of the spherical symmetries explicitly, we can define an angle $\theta_{xy}$ via
\be
L_{xy}=x\frac{\partial}{\partial y}-y\frac{\partial}{\partial x}
\equiv\frac{\partial}{\partial\theta_{xy}},
\ee
while $x,y$ are no longer independent on the sphere,
$\theta_{xy}$ can be arbitrary, as this degree of freedom moves a given point around
on a subset of the spherical surface. From this viewpoint,
the commutator algebra of $L_{xy},L_{yz},L_{zx}$ is obviously unaltered.
More profoundly, it is these rotation generators that fully generate the spherical surface.
Given a particular point in $\mathbb{R}^3$, rotation generators move it around to sweep over
the entire surface of a fixed distance from the origin.

This picture can be exactly generalized to the case of BCFW deformation generators.
We can define an `angle' in a complex spinorial sense for each deformation, via
\be
D_{\<i|j]}\equiv\frac{\partial}{\partial\theta_{\<i|j]}},
\ee
then $\theta_{\<i|j]}$ parameterizes one of the symmetries that preserve momentum conservation,
and hence momentum conservation will not alter the commutator algebra of $D_{\<i|j]}$ at all.
But instead, this constraint is fully generated by $2\,C^2_n=n(n-1)$ BCFW deformation generators.
Given a particular point in $\mathbb{C}^{4n}$, namely the complex spinorial space $(\la_i,\tla_i)$,
BCFW deformation generators move it around to sweep over
the entire codimension-4 surface of a fixed sum of external momenta. And physically, this sum is zero.

Since the commutator algebra is unaltered, we are free to treat all spinors as independent to
derive the commutation relations.
Imagine the 0th step of deformation is $\<i|j]$, then the 1st step can be one of
the four types as named below:
\be
\bal
\<k|l]&=\textrm{independent},\\
\<i|l]\textrm{ or }\<k|j]&=\textrm{straight descendent},\\
\<l|i]\textrm{ or }\<j|k]&=\textrm{skew descendent},\\
\<j|i]&=\textrm{cross descendent}.
\eal
\ee
The generators of first two types commute with that of $\<i|j]$, \ie
\be
\[D_{\<i|j]},D_{\<k|l]}\]=0,~\[D_{\<i|j]},D_{\<i|l]}\]=\[D_{\<i|j]},D_{\<k|j]}\]=0.
\ee
For the last two types,
\be
\bal
\[D_{\<i|j]},D_{\<j|k]}\]=D_{\<i|k]},~\[D_{\<i|j]},D_{\<j|i]}\]=2h_i-2h_j,\\
2h_i=-\la_i^\ap\frac{\partial}{\partial\la_i^\ap}
+\tla_i^{\dot{\ap}}\frac{\partial}{\partial\tla_i^{\dot{\ap}}},~
\[2h_i,D_{\<i|j]}\]=\[D_{\<i|j]},2h_j\]=D_{\<i|j]},
\eal
\ee
where $h_i$ is the helicity operator with respect to the $i$-th particle,
for a function covariant under the little group (an amplitude does have this scaling property).

For the skew descendent case, using the Baker-Campbell-Hausdorff formula
\be
\exp X\exp Y=\exp(X+Y)\exp\(\frac{1}{2}[X,Y]\),~\textrm{for }[X,[X,Y]]=[Y,[X,Y]]=0,
\ee
and due to the commutativity of a straight descendent pair, we have
\be
\[\exp\(z_0D_{\<i|j]}\),\exp\(z_1D_{\<j|k]}\)\]
=\exp\(z_0D_{\<i|j]}+z_1D_{\<j|k]}\)2\sinh\(\frac{1}{2}z_0z_1D_{\<i|k]}\).
\ee
Hence skew descendent deformations $\<i|j]$ and $\<j|k]$ commute if $D_{\<i|k]}$ annihilates
the amplitude, however, this is a too stringent condition which often trivializes the deformations
being used. Therefore in general, skew descendent deformations do not commute.

\subsection{Commutativity at integrand and integral levels}

By applying the BCFW deformation generators,
we perform one constant extraction on rational function $R(\la_i,\tla_i)$
as (different from contours around finite locations, the contour around infinity is clockwise)
\be
CR=\oint_\infty\frac{dz}{z}R(z)=\oint_0\frac{dw}{w}R\(\frac{1}{w}\),
\ee
where the change of variable is $w\equiv1/z$, so the infinity for $z$ is the zero for $w$.
However, the residue at this zero is \textit{not} a naive one. Recall \eqref{eq-10}
in terms of $w$, it reads
\be
R\(\frac{1}{w}\)=\sum_k\frac{(b_{1k}+b_{0k}w)w^{2d_k-1}}{\(a_{2k}+a_{1k}w+a_{0k}w^2\)^{d_k}}
+c_0+\sum_l\frac{c_l}{w^l},
\ee
a naive substitution of $w=0$ will cause divergence in the third term above. On the other hand,
it is clear that after the expansion, the third term actually has no simple pole at $w=0$, since there is
already one $w$ in the denominator of the integrand.
To remove this divergent term, before the contour integration we must Laurent expand $R(1/w)$
around $w=0$, \ie we need to first factor out a divergent factor
$1/w^\gm$ with $\gm\geq1$, leaving a finite fraction at $w=0$, then Taylor expand it around $w=0$.
A simple example is
\be
\frac{b_0+b_1z+b_2z^2}{a_0+a_1z}=\frac{1}{w}\frac{b_2+b_1w+b_0w^2}{a_1+a_0w}
=\frac{b_2}{a_1w}\(1+\frac{b_1}{b_2}w+\frac{b_0}{b_2}w^2\)\(1+\frac{a_0}{a_1}w\)^{-1},
\ee
then we can Taylor expand the finite fraction around $w=0$, and the contour integral
will only pick up the constant part in this expression.
Similarly, performing two successive constant extractions gives
\be
C_1C_0R=\oint_0\frac{dw_1}{w_1}\oint_0\frac{dw_0}{w_0}R\(\frac{1}{w_1},\frac{1}{w_0}\),
\ee
with
\be
R\(\frac{1}{w_1},\frac{1}{w_0}\)=\exp\(\frac{1}{w_1}D_1\)\exp\(\frac{1}{w_0}D_0\)R.
\ee
For independent and straight descendent cases, $[D_1,D_0]=0$, so the order
of deformations is irrelevant, and hence the commutativity of these two types holds at
\textit{integrand} level\footnote{It is possible that two constant extractions commute, even if
they do not commute at integrand level, but we will not consider this trivial case here.}.

However, before performing the integral, to double Laurent expand a fraction is a bit tricky.
First, to properly factor out the overall factor in terms of $w_0$ and $w_1$,
we need to ensure that in
\be
R\(\frac{1}{w_1},\frac{1}{w_0}\)=\frac{1}{w_0^{\gm_0}w_1^{\gm_1}}\frac{P(w_0,w_1)}{\prod_iQ_i(w_0,w_1)},
\ee
both $P(w_0,w_1)$ and $Q_i(w_0,w_1)$ are irreducible polynomials, \ie
there is no common factor $w_0^{\de_0}w_1^{\de_1}$ in each $P_i$ or $Q_i$.
Then, we find the expansions of a fraction in opposite orders
are different, when any of the $Q_i$'s does not contain a constant term.
For example, take $g(w_1,w_0)=1/(w_1+w_0)$ and
expand it around $w_0=0$ (it's impossible to further expand around $w_1=0$), we have
\be
g(w_1,w_0)=\frac{1}{w_1}\(1-\frac{w_0}{w_1}+\frac{w_0^2}{w_1^2}-\ldots\),
\ee
and for the reverse order,
\be
g(w_1,w_0)=\frac{1}{w_0}\(1-\frac{w_1}{w_0}+\frac{w_1^2}{w_0^2}-\ldots\),
\ee
hence they are clearly different. In general, if any of the $Q_i$'s happens to satisfy
$Q_i(0,0)=0$, the double expansion depends on the order. Conversely,
if all $Q_i$'s obey $Q_i(0,0)\neq0$, the order of double expansion is irrelevant.
Since the contour integral picks up the constant part in the expansion only, the
commutativity of the denominator expansion is equivalent to the
commutativity at \textit{integral} level.

In practice, since it is clear that
a detectable pole of merely one of the two successive deformations always contains a constant term
after factoring out a proper factor, we should only focus on the overlap of two sets of detectable poles.
But it's impractical to trace each term at each step, for seeing whether a constant term exists.

Combining everything above, one reaches the conclusion: the
commutativity of $C$ operators holds for independent and straight descendent deformations,
provided that the condition $Q_i(0,0)\neq0$ is satisfied.
These two types may be considered as `good', as they may enjoy the
orthogonal property \eqref{eq-2}. However, merely the good types are not sufficient to capture
all physical pole terms, as will be explained in the end of appendix \ref{app2}.
Hence we will not proceed further in this direction,
but return to seek for the condition of $C_n\cdots C_1C_0=0$, such that
the $(n+1)$ steps can fully capture the amplitude.

Nevertheless, this investigation gives a crucial hint for the subsequent analysis: In the expansion,
the coefficient of $z$ after a deformation becomes a new pole of the corresponding boundary term,
with order one or higher. To be concrete, consider the example below
\be
\frac{1}{a}\to\frac{1}{a+bz}=\frac{w}{(b+aw)}=\frac{w}{b}\(1+\frac{a}{b}w\)^{-1}
=\frac{w}{b}\(1-\frac{a}{b}w+\frac{a^2}{b^2}w^2-\ldots\), \labell{eq-4}
\ee
where $b$ is the only source of poles afterwards.

\section{Systematic Algorithm of Finite, Definite Steps}
\labell{sec3}

In this section,
we propose the systematic process to capture the amplitude after finite, definite steps
of BCFW constant extractions. The condition of correctly completing the calculation
is simply $C_n\cdots C_1C_0=0$. To achieve this,
a form of all poles in the final boundary term is obtained, after a sequence of constant extractions
which is called the `pole concentration'. This sequence is designed for covering \textit{all}
situations so there is no extra restriction such as color order,
and how to optimize it case by case is set aside temporarily.
Having the final form of poles, merely using the information of mass dimension and helicities
is sufficient to judge whether the final boundary term vanishes.

\subsection{Pole concentration}

Now we use pole concentration to capture all poles
regardless of whether they are physical or spurious by applying the logic of \eqref{eq-4}.
Also, each time we perform a BCFW constant extraction
on the amplitude, at least one of its physical poles will be filtered out, and consequently
each corresponding boundary term will contain at least one pole mutated from the original
physical poles.

For example, consider denominator $\<12\>\<23\>$ (the numerator is neglected for our purpose),
under constant extraction $\<1|3]$,
\be
\frac{1}{\<12\>\<23\>}\to\frac{1}{(\<12\>-z\<32\>)\<23\>}\Rightarrow\frac{1}{\<23\>^2},
\ee
where pole $\<12\>$ has been replaced by $\<32\>$. Crucially, under a next constant extraction,
pole $\<23\>^2$ is either unchanged or replaced by another pole of the same order. This means once two poles
are \textit{stacked}, they are stacked forever. The same logic also works for anti-holomorphic poles. For a
multi-particle pole, we first need to turn it into a product of holomorphic and
anti-holomorphic poles, with a proper choice of deformation.
As one example, under constant extraction $\<1|4]$,
\be
\frac{1}{P_{123}^2}\to\frac{1}{P_{123}^2+z\<4|2+3|1]}\Rightarrow\frac{1}{\<4|2+3|1]},
\ee
next, under constant extraction $\<2|5]$,
\be
\frac{1}{\<4|2+3|1]}\to\frac{1}{\<4|2+3|1]-z\<45\>[21]}\Rightarrow\frac{1}{\<45\>[21]},
\ee
then we are again left with two-particle poles.

In general, one can first turn a
multi-particle pole $P^2$ into $\<i_1|P|j_1]$, where $P$ includes either $i_1$ or $j_1$,
note that $p_{i_1}$ or $p_{j_1}$ in $P$ is already filtered out by $\<i_1|$ or $|j_1]$.
Next, one can split $\<i_1|P|j_1]$ by using $\<k|j_2]$ or $\<i_2|l]$,
where $P$ includes $j_2$ or $i_2$ but not $k$ or $l$.
This way turns the pole into a product of one holomorphic and one anti-holomorphic pole.
Then for two-particle poles, once they are stacked, they must mutate as a whole afterwards.
After finite steps, all poles can be encapsulated in only one holomorphic
and one anti-holomorphic pole, with orders larger than one in general.

In appendix \ref{app2}, one sequence of BCFW constant extractions is presented to turn all poles
of the final boundary term into a common denominator, of the expression
given by\footnote{The choice of sequence is not unique,
and how to optimize it to shorten the steps is a very valuable future problem.}
\be
\frac{(\textrm{polynomial})}{\<i_1i_2\>^m[i_3i_4]^{\m}}\times(\textrm{remaining factor}), \labell{eq-5}
\ee
where $i_1,i_2,i_3,i_4$ are four different arbitrary particle labels. The remaining factor
is a rational function, which is dimensionless and helicity-neutral, see \eqref{eq-17} for example.
Note that we \textit{have not} reduced the denominator against the numerator.
At the first glance, the reason to get this final denominator is that we can use one more
deformation, say $\<i_1|i_4]$, to get the maximal large $z$ suppression, since all poles
after concentration are vulnerable to it.
But in fact, there is a less obvious argument for eliminating
the final boundary term without introducing one more step, as will be given later.

Here, $m$ or $\m$ gets contribution from physical holomorphic or anti-holomorphic poles, and both of them get
contributions from physical multi-particle poles. In general, $m$ and $\m$ need not be equal,
since not all possible poles are physical for a particular amplitude.
To see the range of $m,\m$, we will analyze all possible physical poles for various $n$'s.
When $n=4$, only a half of all two-particle poles can appear in the amplitude,
since they are doubly duplicated by
momentum conservation. When $n=5$, there are only two-particle poles, as three-particle poles
are equivalent to them by momentum conservation.
When $n\geq6$, multi-particle poles arise. Their particle numbers range from 3
to $(n-3)$, to avoid duplications of two-particle poles by momentum conservation.
To further avoid duplications of themselves, one can
fix the pole momentum by demanding one pivot particle to be always included,
and then the number of multi-particle poles is reduced by one half.

According to the counting above, the maxima of $m,\m$ are
\be
m_{\max}=\m_{\max}=C_n^2+\frac{1}{2}(C_n^3+\ldots+C_n^{n-3})=2^{n-1}-(n+1),
\ee
which nicely covers the special cases of $n=4$ and $n=5$.

However, there is a little subtlety in \eqref{eq-5}: For a given amplitude,
while $m,\m$ can be easily read off by analyzing all of its non-vanishing factorization limits,
the final boundary term in general contains \textit{not only} poles $\<i_1i_2\>^m[i_3i_4]^{\m}$,
but also the same poles of higher orders from the dimensionless, helicity-neutral remaining factor.
This also occurs in each intermediate step for each corresponding intermediate boundary term.
A simple example is the MHV amplitude $A(1^-,2^-,3^+,4^+)$, given by
\be
A=\frac{\<12\>^3}{\<23\>\<34\>\<41\>},
\ee
and deformation $\<1|3]$ turns it into (recall that $z=1/w$)
\be
A\to\frac{(\<12\>-z\<32\>)^3}{\<23\>\<34\>(\<41\>-z\<43\>)}
=\frac{1}{w^2\<23\>\<34\>}\frac{(-\<32\>+\<12\>w)^3}{-\<43\>+\<41\>w}
=\frac{1}{w^2\<23\>\<34\>}\frac{\<32\>^3}{\<43\>}
\(1-\frac{\<12\>}{\<32\>}w\)^3\(1-\frac{\<41\>}{\<43\>}w\)^{-1},
\ee
note that pole $\<41\>$ is turned into $\<43\>$, but its order can be larger than one.
Explicitly, the corresponding boundary term is
\be
\bal
C_{\<1|3]}A&=\frac{1}{\<23\>\<34\>\<43\>}\(\frac{\<41\>^2\<32\>^3}{\<43\>^2}
-3\frac{\<41\>\<12\>\<32\>^2}{\<43\>}+3\<12\>^2\<32\>\)\\
&=-\frac{\<12\>^2\<32\>}{\<23\>\<34\>^2}\(\frac{\<41\>^2\<32\>^2}{\<43\>^2\<12\>^2}
-3\frac{\<41\>\<32\>}{\<43\>\<12\>}+3\), \labell{eq-17}
\eal
\ee
where the term in parentheses is the remaining factor of \eqref{eq-5}. The advantage of packing up
many pole terms into a dimensionless helicity-neutral factor is that,
if we can show this representative factor cannot exist, the full expression including terms with
higher-order poles must also be forbidden.

One digressive comment is that, so far we have found BCFW deformations to be the only type which
admits a feasible pole concentration. A counterexample is, there is no straightforward pole concentration
for Risager deformations \cite{Risager:2005vk}.
We will not further explain the claim here but it is not hard to confirm it.
This is another specialty of BCFW deformations, in addition to that
BCFW deformations automatically preserve (or generate) the momentum conservation constraint.

\subsection{Kinematic mass dimension}

To prepare for the later analysis, we will study the general information of an amplitude:
mass dimension and helicities, with which
the applicable range of multi-step BCFW recursion relations can be clarified.

First, for QFTs in 4-dimension, the mass dimension of an $n$-particle amplitude is $(4-n)$.
We can use the LSZ reduction formula to prove this.
Schematically, an $n$-particle amplitude $A$ is defined via
\be
\prod^n\(\int d^4x\,e^{ipx}\varepsilon\Delta\)\<\Phi_1\ldots\Phi_n\>=\delta^4\(\sum p\)A,
\ee
where $\<\Phi_1\ldots\Phi_n\>$ is the $n$-point function, $\varepsilon$ and $\Delta$ are
the wave-function and kinematic operator for each field $\Phi$.
For a bosonic field, the mass dimensions of $\varepsilon$, $\Delta$ and $\Phi$ are 0, 2 and 1 respectively,
for a fermionic field, the mass dimensions of $\varepsilon$, $\Delta$ and $\Phi$ are $1/2$, 1 and $3/2$
respectively. Hence the mass dimension of
\be
\int d^4x\,e^{ipx}\varepsilon\Delta\Phi
\ee
is $-1$. There are $n$ such pieces, plus the momentum conserving delta function,
the mass dimension of $A$ is clearly $(4-n)$.

One special bosonic field is the graviton. By the perturbative definition
$g_{\mu\nu}=\eta_{\mu\nu}+h_{\mu\nu}$, it should be dimensionless. To treat it as ordinary bosonic fields,
we need to redefine it via
\be
g_{\mu\nu}=\eta_{\mu\nu}+\kappa\,h_{\mu\nu},
\ee
such that $h_{\mu\nu}$ carries mass dimension 1 and $\kappa$ carries $-1$. Choosing $\kappa$ to be
$\sqrt{8\pi G}$, the free field part of Einstein-Hilbert Lagrangian takes an analogous form as
those of ordinary bosonic fields. Consequently, $\kappa$ becomes the coupling constant of gravity.

One can also rediscover $\kappa$ via the on-shell method.
For gravity, three-particle amplitudes including at least one graviton are
\be
\bal
&A(1^{-h},2^{+h},3^{-2})=\kappa\,\<12\>^{-2}\<23\>^{-2h+2}\<31\>^{+2h+2},\\
&A(1^{-h},2^{+h},3^{+2})=\kappa\,[12]^{-2}[23]^{+2h+2}[31]^{-2h+2},
\eal
\ee
where $h=0,1/2,1,2$, for all `realistic' theories. No matter which value $h$ takes, $\kappa$ always carries
mass dimension $-1$, since the mass dimension of three-particle amplitudes is 1.

In general, we can reverse the logic above and define the `kinematic mass dimension' of an $n$-particle
amplitude as
\be
D=4-n-\sum_i(D_c)_i, \labell{eq-7}
\ee
where $(D_c)_i$ is the mass dimension of the coupling constant for each vertex.
From now on, we will focus on the kinematic part of an amplitude which is a function of spinorial products.

For the Standard Model\footnote{Strictly speaking, we mean all Standard-Model-type interactions
of massless particles, and only dimensionless couplings are involved.},
all coupling constants are dimensionless so $D=4-n$,
which is a non-positive number for $n\geq4$.
For gravitational interactions $D_c=-1$ and we will show that $D=2$.

When $D<0$, there is at least one irreducible denominator of the amplitude,
which means a pole to be detected by BCFW deformations. Conversely, when $D\geq0$,
the amplitude may admit some invulnerable terms to BCFW deformations which include polynomials and
`pseudo polynomials'. The classification of these objects can be found in appendix \ref{app3}.

\subsection{The master formula}

By using mass dimension and helicities,
we now derive the master formula for the subsequent discussions. After pole concentration,
the final boundary term schematically reads\footnote{In general, the final boundary term's numerator is a
polynomial but we only focus on one term, as the identical analysis applies to all terms.
The remaining factor of \eqref{eq-5} is dropped, since it does not contribute to the master formula.}
\be
\frac{1}{\<12\>^m[34]^{\m}}\prod_{i=1}^n\<i|^{\ap_i}\prod_{i=1}^n[i|^{\bt_i}, \labell{eq-13}
\ee
where we have temporarily taken $i_{1,2,3,4}=1,2,3,4$.
The reason to use un-contracted spinors is that, it is more compact to capture the helicity information,
and it can save the Schouten identity manipulations, as one can freely recombine them to get the desired
spinorial products. Of course, the cost is that one needs to rule out all those illegitimate combinations.
This treatment is similar to the methods used in \cite{Cohen:2010mi,McGady:2013sga}.
Now the helicity configuration enforces that
\be
\bal
-2h_1+m&=\ap_1-\bt_1,\\
-2h_2+m&=\ap_2-\bt_2,\\
2h_3+\m&=\bt_3-\ap_3,\\
2h_4+\m&=\bt_4-\ap_4,\\
2h_i&=\bt_i-\ap_i,~(i=5,\ldots,n)\\
\eal
\ee
where $m,\m$ are known for a particular amplitude. Note that there are $2n$ variables, with only $n$
helicity constraints. We will fully exploit the $n$ remaining degrees of freedom to derive the master formula.

The kinematic mass dimension of \eqref{eq-13} is
\be
D'=-(m+\m)+\frac{1}{2}\(\sum_{i=1}^n\ap_i+\sum_{i=1}^n\bt_i\),
\ee
where obviously, $\sum\ap$ and $\sum\bt$ must be both even to form spinorial products.
Also, we have $m+\m\geq1$ with $m,\m\geq0$, and $\ap,\bt\geq0$. For a legitimate final boundary term,
$D'$ equals to $D$ defined in \eqref{eq-7}.

When $D\neq D'$ under all circumstances, the correct dimension and helicities cannot be
satisfied simultaneously, and then the final boundary term is eliminated. One direct way to achieve this
inconsistency is to show $D'_{\min}$ is larger than $D$. First we need to figure out this minimum
by eliminating one variable for each particle, as
there are two variables $\ap_i$ and $\bt_i$ to be chosen. For $i=5,\ldots,n$,
\be
\frac{1}{2}(\ap_i+\bt_i)=-h_i+\bt_i=h_i+\ap_i,
\ee
when $h_i$ is negative, $(-h_i+\bt_i)$ is guaranteed to be positive, similarly
when $h_i$ is non-negative, $(h_i+\ap_i)$ is guaranteed to be non-negative.
To manifest the non-negativity of $D'$, our choice is
\be
\frac{1}{2}(\ap_i+\bt_i)=|h_i|+\min(\ap_i,\bt_i).
\ee
Extending this logic for all particles, yields
\be
D'=-(m+\m)+\sum_{i=1,2}\(\lv h_i-\frac{m}{2}\rv+\min(\ap_i,\bt_i)\)
+\sum_{i=3,4}\(\lv h_i+\frac{\m}{2}\rv+\min(\ap_i,\bt_i)\)
+\sum_{i=5}^n(|h_i|+\min(\ap_i,\bt_i)), \labell{eq-6}
\ee
which is the \textit{master formula}, and explicitly,
\be
\sum_{i=5}^n(|h_i|+\min(\ap_i,\bt_i))=\sum_{h<0}(-h_i+\bt_i)+\sum_{h\geq0}(h_i+\ap_i),
\ee
which separates the sum into two parts according to the helicities. The final boundary term
\eqref{eq-13} now reads ($p_i=|i\>[i|$ is a helicity-neutral momentum
with additional mass dimension 1)
\be
\frac{1}{\<12\>^m[34]^{\m}}\prod_{i=1,2}\([i|^{2h_i-m}p_i^{\ap_i}\Big/\<i|^{-2h_i+m}p_i^{\bt_i}\)
\prod_{i=3,4}\([i|^{2h_i+\m}p_i^{\ap_i}\Big/\<i|^{-2h_i-\m}p_i^{\bt_i}\)
\prod_{h<0}\<i|^{-2h_i}p_i^{\bt_i}\prod_{h\geq0}[i|^{2h_i}p_i^{\ap_i},
\ee
where / means one of two candidate expressions is chosen to manifest the non-negativity of $D'$,
as this choice also manifests the `extra neutral momenta', in addition to the `net spinors' that carry
the helicity information. While the latter content is mandatory, the former is optional since it is brought
in to fill the extra capacity of mass dimension. There is no unique choice of picking these extra $\ap$'s and
$\bt$'s as long as the total dimension is correct.

For $i=5,\ldots,n$, $|h_i|$ is trivially non-negative.
For $h_1,h_2,h_3,h_4$, careful analysis is needed as $m,\m$ are involved. Rewrite \eqref{eq-6} as
\be
D'=-(m+\m)+T_{1234}+\sum_{i=1}^n|h_i|+\sum_{i=1}^n\min(\ap_i,\bt_i), \labell{eq-8}
\ee
where
\be
T_{1234}\equiv
\sum_{i=1,2}\(\lv h_i-\frac{m}{2}\rv-|h_i|\)+\sum_{i=3,4}\(\lv h_i+\frac{\m}{2}\rv-|h_i|\),
\ee
since $m,\m$ and $\sum|h|$ are fixed, we only need to manipulate $T_{1234}$.
It's easy to check that
\be
\bal
\lv h_i-\frac{m}{2}\rv-|h_i|&=\left\{\begin{array}{ll}
m/2, & h_i<0\\
m/2-2h_i, & 0\leq h_i<m/2\\
-m/2, & m/2\leq h_i
\end{array}\right.\\
\lv h_i+\frac{\m}{2}\rv-|h_i|&=\left\{\begin{array}{ll}
-\m/2, & h_i<-\m/2\\
\m/2+2h_i, & -\m/2\leq h_i<0\\
\m/2, & 0\leq h_i
\end{array}\right.
\eal
\ee
to maximize $T_{1234}$, one must take $h_1,h_2$ to be two of the smallest helicities,
and $h_3,h_4$ to be two of the largest helicities in the process\footnote{Now we should strictly use
$i_1,i_2,i_3,i_4$ instead of $1,2,3,4$ to admit a possible relabeling,
when we implement the desired arrangement of pole concentration.}.

On the other hand, even if one chooses $h_1,h_2,h_3,h_4$ arbitrarily among all $h_i$'s,
$D'_{\min}$ is \textit{no less than zero} after a similar relabeling.
To see this, let's rewrite \eqref{eq-6} as
\be
D'=-(m+\m)+T'_{1234}+\sum_{i=5}^n|h_i|+\sum_{i=1}^n\min(\ap_i,\bt_i),
\ee
and focus on the quantity
\be
T'_{1234}\equiv\sum_{i=1,2}\lv h_i-\frac{m}{2}\rv+\sum_{i=3,4}\lv h_i+\frac{\m}{2}\rv,
\ee
by this definition $T'_{1234}$ has a simple geometric meaning:
It is the sum of the distances of four line segments stretching from $h_1,h_2$ to line $h=m/2$, and
from $h_3,h_4$ to $h=-\m/2$, as shown in Figure \ref{fig-0}.
It's easy to find that its minimum is $(m+\m)$, when four points
are on one horizontal line and the line is within the region between $h=m/2$ and $h=-\m/2$.
When this horizontal line moves outside the region, $T'_{1234}$ increases by
$4\times(\textrm{distance above or below the region})$.
When this line is not horizontal, one can see that $T'_{1234}$
is always larger than $(m+\m)$ after a relabeling such that $h_1,h_2\leq h_3,h_4$.
Since $T'_{1234}$ is no less than $(m+\m)$, $D'_{\min}$ is always non-negative.

\begin{figure}
\begin{center}
\includegraphics[width=0.6\textwidth]{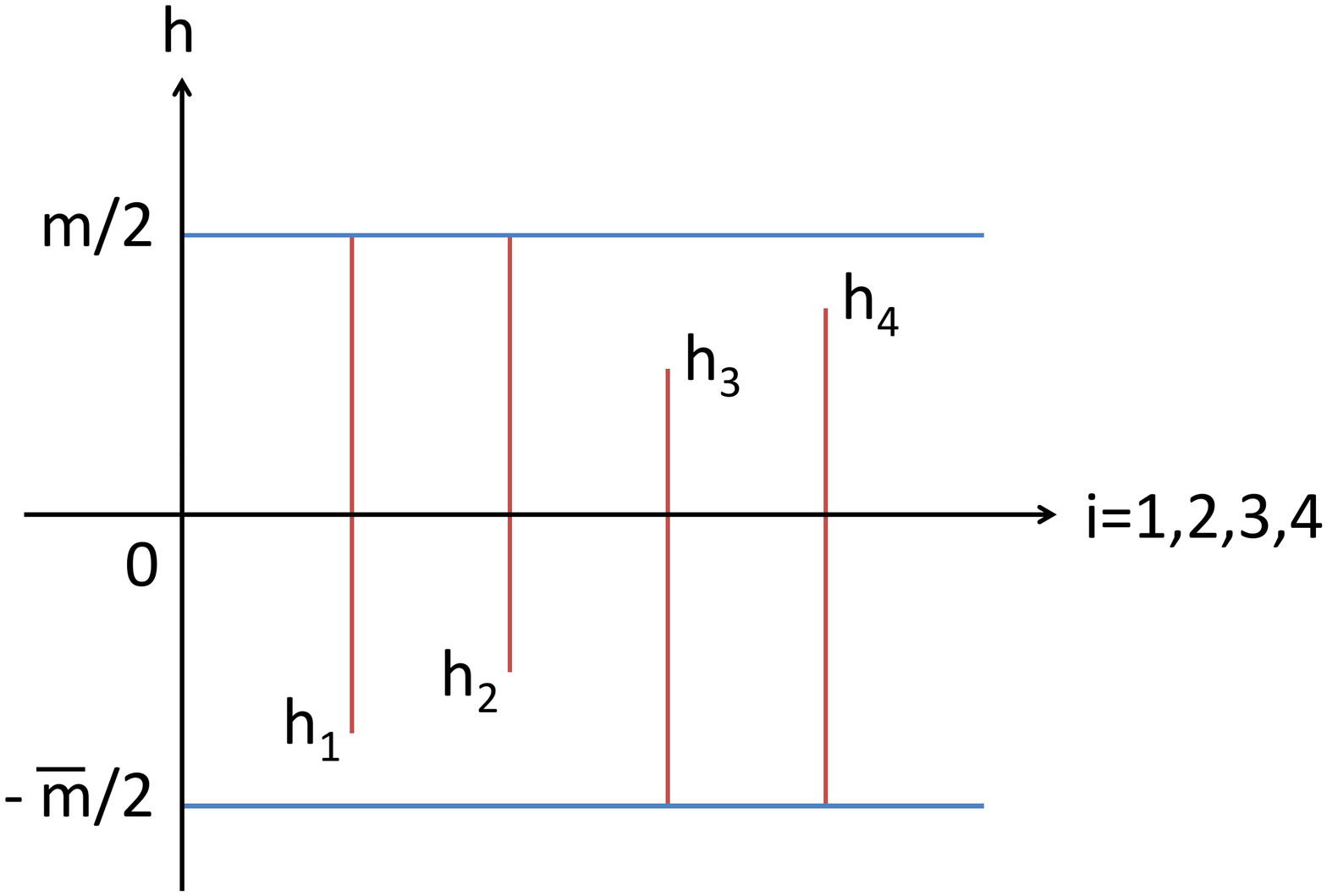}
\caption{Sum of the distances of four line segments. It is obviously no less than $(m+\m)$.} \labell{fig-0}
\end{center}
\end{figure}

Also note that
\be
\bal
-\frac{m}{2}+\lv h_i-\frac{m}{2}\rv-|h_i|&=0\Big/(-2h_i)\Big/(-m),\\
-\frac{\m}{2}+\lv h_i+\frac{\m}{2}\rv-|h_i|&=(-\m)\Big/(2h_i)\Big/0,
\eal
\ee
which always take integer values. According to \eqref{eq-8}, if $\sum|h|$ is fractional,
$D'$ must be fractional.

\subsection{Inconsistency elimination}

We are prepared to show that all massless tree amplitudes except those admitting
(pseudo) polynomials of given mass dimension and helicities, shall be fully determined by multi-step BCFW
recursion relations. In the following analysis, we pretend not to know any knowledge of QFT
in the Lagrangian paradigm besides mass dimension and helicities. These are the only data needed to
construct three-particle amplitudes, and recursions extend them to all higher-point ones.

First note that,
after pole concentration, there is no need to find any further deformation to kill the final boundary term,
because another choice will not change its pole form, so one can always rearrange the entire sequence
of pole concentration to reach the desired relabeling. Hence it must be the last step if chosen properly,
and the direct way to use is the `inconsistency elimination'.

Within its own framework, if inconsistency elimination can exclude the final boundary term,
the new algorithm should be a completely independent approach for calculating a particular amplitude.
Otherwise, terms that survive it need to be identified, similar to (pseudo) polynomials.
In fact, we do discover a new type of object called the `saturated fraction' in this way.

From the master formula \eqref{eq-6}, it is already known that $D'_{\min}\geq0$. Therefore if $D<0$,
$D<D'_{\min}$ always holds. This tells us the nontrivial cases are of $D\geq0$.
When $D\geq D'_{\min}$,
we have the inconsistency criteria below to further eliminate the final boundary term:\\

(1) Fractional Dimension (FD): If $D'=\textrm{fractional}$.
This arises when $\sum|h|=\textrm{fractional}$, but Lorentz invariance demands the dimension of
\eqref{eq-13} to be an integer. This also implies that fermions must appear
in pairs to be consistent.\\

(2) Pair Mismatch (PM): If $\sum\ap=\textrm{odd}$ and $\sum\bt=\textrm{odd}$,
when FD is excluded already. In this case,
\eqref{eq-13} cannot be written as a fraction in terms of Lorentz invariant spinorial products,
even though the dimension of \eqref{eq-13} is an integer.\\

(3) Spinor Excess (SE): If there exists an $i$, such that
$\ap_i>\sum_{j\neq i}\ap_j$ or $\bt_i>\sum_{j\neq i}\bt_j$. In this case, spinorial contraction
will force \eqref{eq-13} to vanish, even though $\sum\ap=\textrm{even}$ and $\sum\bt=\textrm{even}$.\\

Altogether, there are four layers of inconsistency criteria:

(0) $D<D'_{\min}$.

(1) If $D\geq D'_{\min}$, consider FD.

(2) If $D\geq D'_{\min}$, and FD is excluded, consider PM.

(3) If $D\geq D'_{\min}$, and both FD and PM are excluded, consider SE.\\

It's obvious that these inconsistency criteria only mention general properties of field theories.
Hence inconsistency elimination is theory independent, while in practice
knowing some theory dependent properties would help simplify discussions case by case.
If the final boundary term can survive all four criteria, it must
admit `saturated fractions' (SF). Note that we have already set aside (pseudo) polynomials,
because they can be identified \textit{without} using the master formula \eqref{eq-6}.
Altogether, there are three types of objects invulnerable to BCFW deformations:

(1) Polynomials, such as $\<12\>$.

(2) Pseudo polynomials of $n=4$, such as $[34]/\<12\>$. When $\<12\>\to0$, we must also have $[34]\to0$
since we are using a BCFW deformation. Then the ratio $[34]/\<12\>$ is finite like a polynomial.

(3) Saturated fractions of $n\geq5$, such as $[34][56]/\<12\>$ of $n=6$. When $\<12\>\to0$, the fraction
becomes divergent, so it is different from pseudo polynomials. But somehow similarly, any BCFW deformation
of $n=6$ fails to render such a fraction vanish under the large $z$ limit.

Among these three objects,
a polynomial is completely inert to BCFW constant extractions, in fact it is invulnerable to
any type of deformation in on-shell methods.
A pseudo polynomial is also completely inert to BCFW constant extractions,
while this requires momentum conservation. A saturated fraction is form-inert to BCFW
constant extractions\footnote{Here we mean a pure saturated fraction.
A mixed saturated fraction is a pure saturated fraction times a polynomial.
Its transform under a constant extraction will be demonstrated in appendix \ref{app3}.},
but with particle labels rearranged. The latter two objects are vulnerable to
other types of deformations, such as Risager deformations \cite{Risager:2005vk}.
The detailed exploration of all these three types is presented in appendix \ref{app3}.

So far, we have witnessed how the systematic process of multi-step BCFW recursion relations
can be arranged to determine a particular amplitude. In summary, there are four steps:

(1) Analyze all non-vanishing factorization limits to determine the amplitude's common denominator,
which is a product of all physical poles. This stage can be done almost purely diagrammatically.

(2) Figure out the amplitude's kinematic mass dimension, then combine this with its helicity
configuration to identify all possible (pseudo) polynomials.
If none of them arises, we assume the amplitude can be fully determined and proceed to the next step.

(3) Choose four particle labels (two of the smallest helicities and two of the largest ones)
to determine the denominator's form of the final boundary term,
and arrange a sequence of BCFW constant extractions to proceed pole concentration.
This sequence must be able to capture all physical poles, such that each of them contributes to the final
denominator via powers $m,\m$. Ensuring this, the sequence should be as concise as possible.
Such an optimization is a very valuable future problem.

(4) Use all four inconsistency criteria layer by layer
to eliminate the final boundary term. If it fails,
identify all possible saturated fractions. Then discuss whether these
saturated fractions are legitimate, if not, clarify the
argument to rule them out as described in the end of appendix \ref{app3}.
This delicate treatment to remove all dependence on spurious poles
is another valuable future problem.

\section{Applications in Standard Model plus Gravity}
\labell{sec4}

Knowing the general guide of multi-step BCFW recursion relations, naturally we would like to see
how it applies to specific theories. As familiar examples, realistic theories,
\ie the Standard Model plus gravity\footnote{Any massless theory can be analyzed analogously.
However, as the amplitude's kinematic mass dimension goes up, more types of
(pseudo) polynomials and saturated fractions may arise,
and one needs to identify them carefully.} are being considered.
For simplicity, we will assume that all particles are massless.

For reader's convenience, we rewrite the master formula below
\be
D'=-(m+\m)+\sum_{i=1,2}\lv h_i-\frac{m}{2}\rv+\sum_{i=3,4}\lv h_i+\frac{\m}{2}\rv
+\sum_{i=5}^n|h_i|+\sum_{i=1}^n\min(\ap_i,\bt_i), \labell{eq-11}
\ee
recall that one should take $h_1,h_2$ to be two of the smallest helicities,
and $h_3,h_4$ to be two of the largest helicities in the process.

\subsection{Two separated sectors}

Let's first consider Standard Model and gravity separately. For the Standard Model,
the previous section gives $D=4-n$.
On the other hand, since $D'_{\min}$ is no less than zero,
when $D\leq-1$, the final boundary term must be eliminated.
This directly tells that all Standard Model amplitudes of $n\geq5$ are solvable,
leaving amplitudes of $n=4$.
The $n=4$ case admits (pseudo) polynomials 1, $(\<34\>/[12])^{\pm1}$ and $(\<34\>/[12])^{\pm2}$,
but the last one will be excluded. As it is proved in the next subsection, any
amplitude's helicity configuration in gauge theory must be between MHV and anti-MHV.

For pure gravity, assuming that one of the Feynman diagrams of
an $n$-particle amplitude contains $v_m$ $m$-point vertices and $p$ internal propagators, it is clear that
\be
\sum_mm\,v_m-2p=n,~\sum_mv_m=p+1,~\Longrightarrow~\sum_m(m-2)v_m=n-2,
\ee
and each $m$-point vertex brings in $(m-2)$ $\kappa$'s, hence from \eqref{eq-7} we have
\be
D=4-n-(-1)\sum_m(m-2)v_m=4-n+(n-2)=2,
\ee
now we compare $D$ with $D'$. When $n\geq6$, $D'_{\min}=4>2$ so this is
completely solvable. When $n=5$, from \eqref{eq-11}, and in the most conservative situation
$-\m/2\leq-2<2\leq m/2$, the all-plus helicity configuration (similar for all-minus) admits
the saturated fraction
\be
\frac{[34]^2[35]^2[45]^2}{\<12\>^4}. \labell{eq-16}
\ee
When $n=4$, the all-plus helicity configuration admits pseudo polynomials such as
\be
\frac{[34]^4}{\<12\>^4}P^2_{xy},~\frac{[34]^5}{\<12\>^5}\<13\>\<24\>, \labell{eq-18}
\ee
where $x,y$ are unspecified, while the all-but-one-minus case gives $D'_{\min}=4>2$ already.

However, similar to gauge theory, any amplitude's helicity configuration
in pure gravity must also be between MHV and anti-MHV.
And for the MHV configuration, \eqref{eq-11} gives $D'_{\min}=8>2$.
Therefore pure gravity is completely solvable.

In general, for Standard Model plus gravity we have $D\leq2$.
Note that an amplitude which contains gravitational vertices only always obeys $D=2$,
regardless of how many or what kinds of external legs it owns, as later shown by \eqref{eq-12}.
This specialty implies that one can arbitrarily attach more particles to a known amplitude
without changing its mass dimension, via gravitational interactions.

\subsection{MHV configuration of gauge theory and gravity}

For the usual gauge (Yang-Mills) theory and gravity, based on the knowledge of multi-step BCFW recursion
relations, one can prove that the helicity configuration of any non-vanishing
amplitude must be between MHV and anti-MHV, \ie there is no all-plus or all-but-one-minus configuration.

From Lorentz invariance and little group scaling, the three-particle amplitudes
for gauge theory and gravity are known to be $(s=1,2)$
\be
A(1^{-s}\,2^{-s}\,3^{+s})=g_{s,--+}\(\frac{\<12\>^3}{\<23\>\<31\>}\)^s,~
A(1^{+s}\,2^{+s}\,3^{-s})=g_{s,++-}\(\frac{[12]^3}{[23][31]}\)^s,
\ee
while the $F^3$-type $(s=1)$ or the $R^3$-type $(s=2)$ three-particle amplitudes are
\be
A(1^{-s}\,2^{-s}\,3^{-s})=g_{s,---}\(\<12\>\<23\>\<31\>\)^s,~
A(1^{+s}\,2^{+s}\,3^{+s})=g_{s,+++}\([12][23][31]\)^s.
\ee
For gauge theory the coupling constant is dimensionless and for gravity the coupling constant carries mass
dimension $-1$, while mass dimensions of $g_{s,---}$ and $g_{s,+++}$ are $-2$ for $s=1$ and $-5$ for $s=2$,
so the all-plus and all-minus three-particle amplitudes are excluded.

To exclude the all-plus and all-but-one-minus amplitudes, we need to show that these configurations
have vanishing contributions from either (BCFW detectable) factorization limits, or (BCFW undetectable)
invulnerable objects which include (pseudo) polynomials and saturated fractions. Recall the key identity of
multi-step BCFW recursion relations
\be
I=P_n+C_nP_{n-1}+\ldots+C_nC_{n-1}\cdots
C_2P_1+C_nC_{n-1}\cdots C_2C_1P_0+C_nC_{n-1}\cdots C_2C_1C_0,
\ee
the terms with $P_i$'s and the term
$C_n\cdots C_1C_0$ represent these two parts of contributions respectively, so when they both vanish,
the corresponding amplitude must vanish.

For factorization limits,
an inductive observation is: An all-plus amplitude can only factorize into one lower-point all-plus amplitude
and one all-but-one-minus amplitude. An all-but-one-minus amplitude can only factorize into two lower-point
all-but-one-minus amplitudes, or one lower-point all-plus amplitude and one MHV amplitude.

For the all-plus case, we will finally recurse down to $A(+++)$, so its factorization limit is zero.
This also excludes the second possibility of factorization for the all-but-one-minus amplitude, then for the latter we will finally recurse down to $A(+++-)$, which can further factorize into two $A(++-)$'s superficially.
However, when we use a BCFW deformation to calculate this part, its contribution is zero.
This is due to the fact that both of its sub-amplitudes
are anti-holomorphic functions, while the non-vanishing BCFW construction requires one to
be holomorphic and the other anti-holomorphic.

Therefore (BCFW detectable) factorization limits of the all-plus and all-but-one-minus configurations
are zero, then we consider (pseudo) polynomials and saturated fractions. Note the previous discussion
has not separated gauge theory and gravity yet, hence it holds for both.

For gauge theory, the only invulnerable object is the pseudo polynomial of $n=4$,
as the reader may check this in appendix \ref{app3}, given by
\be
\frac{[34]^2}{\<12\>^2},
\ee
which is related to $A(++++)$. However, when one uses the holomorphic factorization limit (which is the only
type of effective deformation to detect pole $\<12\>^2$, such as Risager deformation as a familiar example)
to send $\la_1\to\la_2$, this pole is spurious since its order is two,
which excludes the pseudo polynomial above.

For gravity, the only invulnerable objects are saturated fraction \eqref{eq-16} related to
$A(+++++)$, and pseudo polynomial \eqref{eq-18} related to $A(++++)$.
Since both of them contain spurious poles, the argument above can also excluded them.
Now the proof is done.

We would like to emphasize that here we have used general information only, namely mass dimension,
helicities and factorization limits, without any further aid such as supersymmetry.

\subsection{Simplified diagrammatic rules}

To simplify the general discussion of Standard Model plus gravity,
we introduce the diagrammatic rules called `stretch and shrink'.
The first example is the gauge interaction, as shown in Figure \ref{fig-1}.

\begin{figure}
\begin{center}
\includegraphics[width=0.4\textwidth]{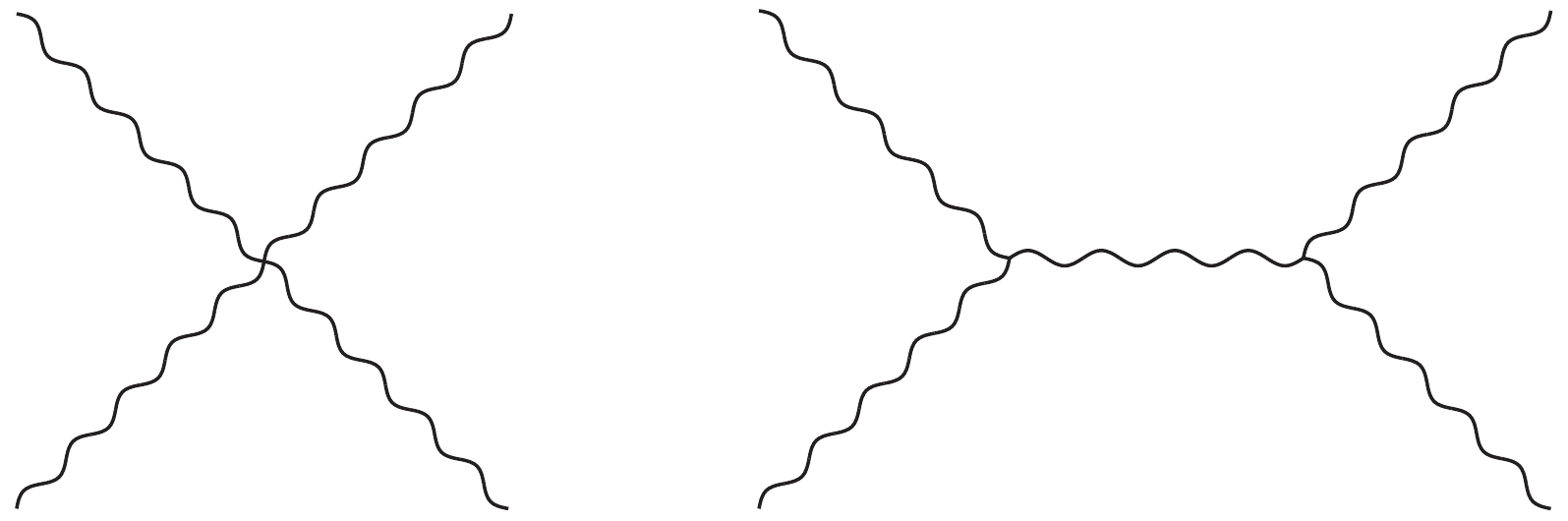}
\caption{4-point gauge vertex is `equivalent' to two connected 3-point vertices.
Wavy lines represent gauge bosons.} \labell{fig-1}
\end{center}
\end{figure}

Fixing four external gauge bosons, this 4-point vertex can be `stretched' into
two connected 3-point vertices, without changing the vertex's mass dimension.
This tricky equivalence holds at the level of mass dimension and helicities, which are the
only information required for inconsistency elimination.

In other words, we have chosen a representative
sub-diagram to encode the same information of mass dimension and helicities, and reduce the types of
equivalent sub-diagrams in the analysis. Following this logic,
all higher-point vertices in Standard Model plus gravity can be stretched into a number of
connected 3-point vertices, except the special $\phi^4$ vertex.
This simplified rule is notably advantageous in gravitational interactions.

\begin{figure}
\begin{center}
\includegraphics[width=0.5\textwidth]{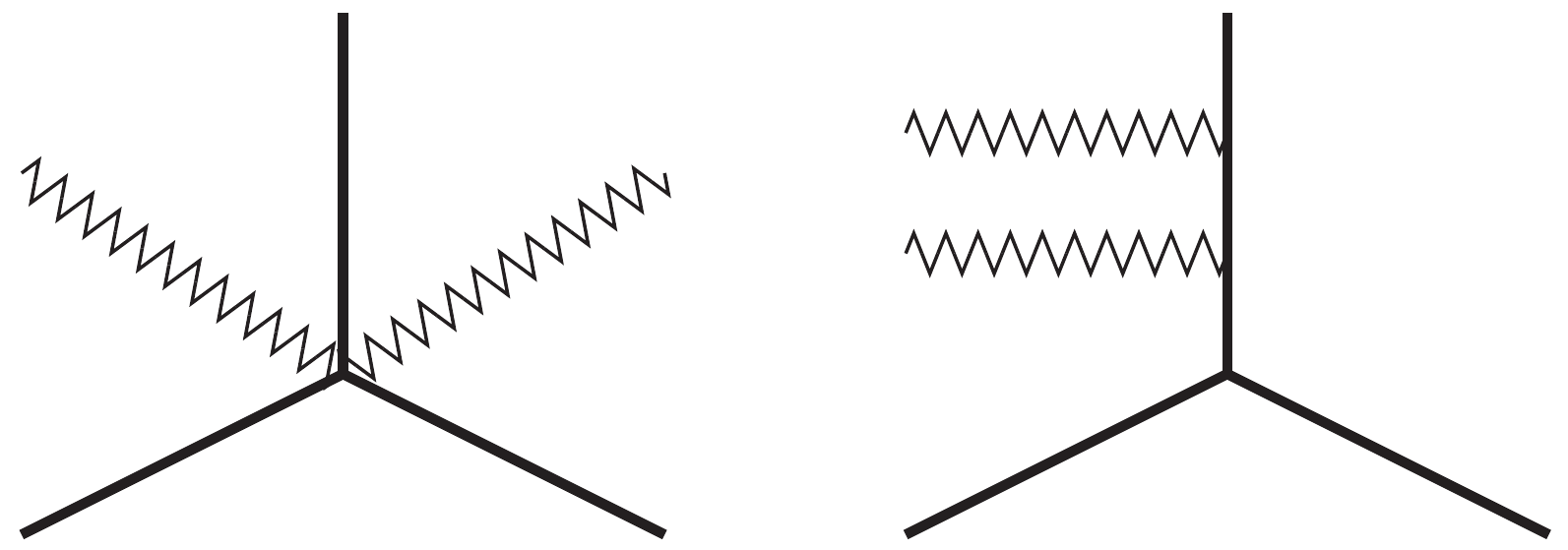}
\caption{Stretch or shift rule of a gravitational vertex. Bold lines represent
Standard Model particles, and zigzag lines represent gravitons.} \labell{fig-2a}
\end{center}
\end{figure}

As shown in Figure \ref{fig-2a}, gravitons can be shifted from any place to any place in a sub-diagram,
without changing its mass dimension. Physically, this is because gravity is universal, gravitons can emit
from any part of a system. Mathematically, this is because an $m$-point gravitational vertex carries
coupling constant $\kappa^{m-2}$, where $\kappa$ carries mass dimension $-1$.
This vertex can contain gravitons only or have Standard Model lines attached to it.
Therefore, the $m$-point vertex can be stretched into $(m-2)$ connected 3-point vertices,
with the exception of $\phi^4$ vertex.

For convenience we define the `gravitational component', as shown in Figure \ref{fig-2b}.
All vertices within this component are gravitational, while its external legs can be either
Standard Model particles or gravitons, or both. There is one special graviton which will attach to another
component. A trivial case is that there is no vertex at all, so this special graviton becomes
the only component.

\begin{figure}
\begin{center}
\includegraphics[width=0.3\textwidth]{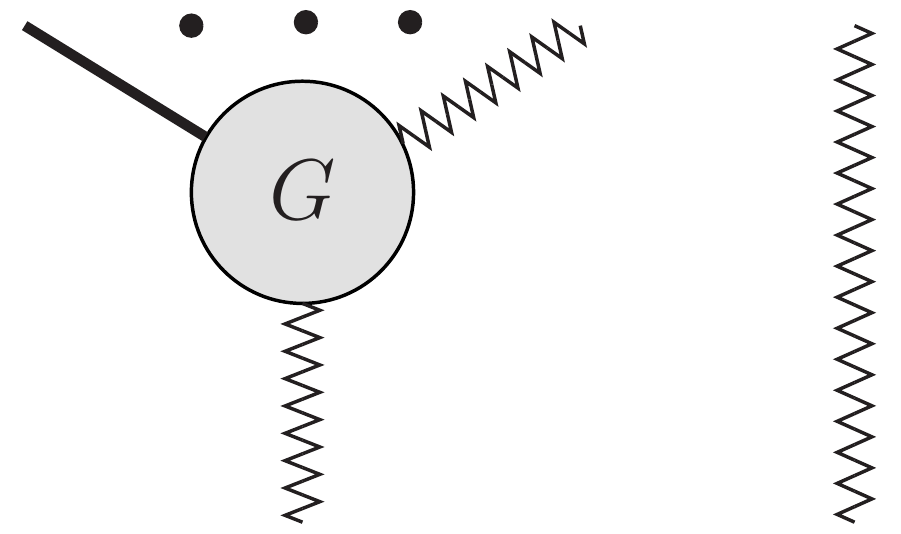}
\caption{Gravitational component and its trivial case.
The unspecified lines between two external lines can be of either type of these two specified lines `on the boundary'. In this case, they can be either Standard Model particles or gravitons, or a mixture of both.} \labell{fig-2b}
\end{center}
\end{figure}

Gravitational components also obey the simplified rules, and for convenience they are usually
shrunk into one component, as shown in Figure \ref{fig-2c}. This pack-up can reduce many sub-diagrams of
gravitational components to one sub-diagram. It's free to attach (or detach) a gravitational component
to (or from) another component, since this operation will not change its mass dimension.

\begin{figure}
\begin{center}
\includegraphics[width=1.0\textwidth]{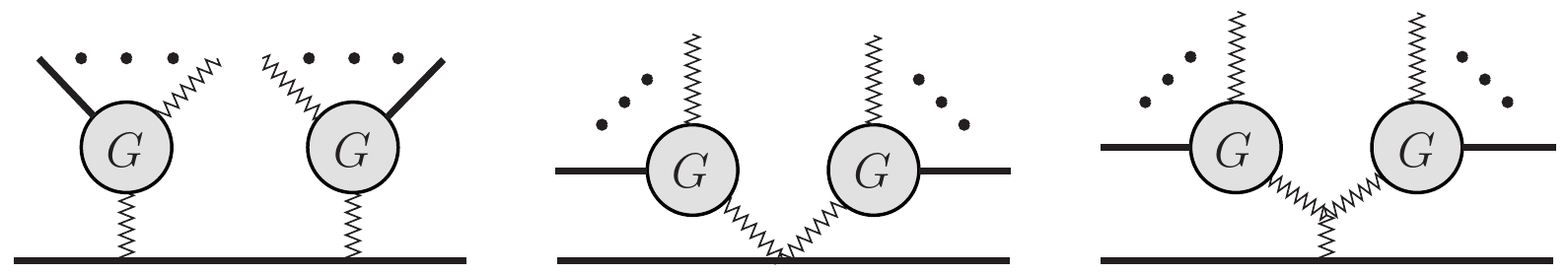}
\caption{Gravitational components can be packed into one component.} \labell{fig-2c}
\end{center}
\end{figure}

Summarizing the simplified diagrammatic rules, we are now left with the representative vertices only,
as shown in Figure \ref{fig-3}.

\begin{figure}
\begin{center}
\includegraphics[width=0.8\textwidth]{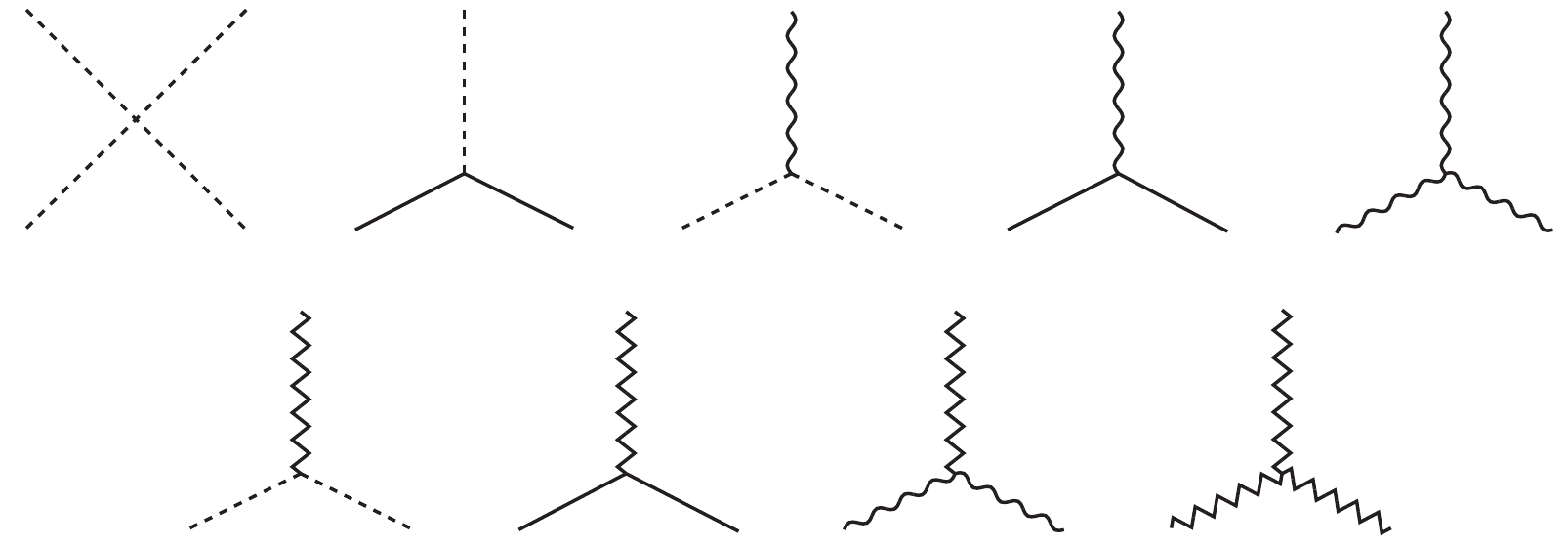}
\caption{All representative vertices in Standard Model plus gravity. Solid lines represent fermions,
and dashed lines represent scalars.} \labell{fig-3}
\end{center}
\end{figure}

\subsection{Amplitudes and (pseudo) polynomials of $D=0,1,2$}

In Standard Model plus gravity, the nontrivial cases are $D=0,1,2$.
By using the representative vertices in Figure \ref{fig-3},
the following discussion is considerably shorten.

$\mathbf{D=0}$ \textbf{case}:
First we consider $D=0$, all possible amplitudes are listed diagrammatically in Figure \ref{fig-4a}.
Similar to gravitational components, Standard Model components presented here only contain Standard Model
vertices. Also note the Standard Model components attached by a single graviton
and a nontrivial gravitational component, are listed \textit{separately} for clarity.

For $D'_{\min}=0$, (pseudo) polynomials arise when all helicities are the same. Now the last three
diagrams in Figure \ref{fig-4a} are excluded, since a three-point Standard Model vertex can never have three
same helicities. The second diagram is also excluded since a graviton has helicity $\pm2$.

\begin{figure}
\begin{center}
\includegraphics[width=0.75\textwidth]{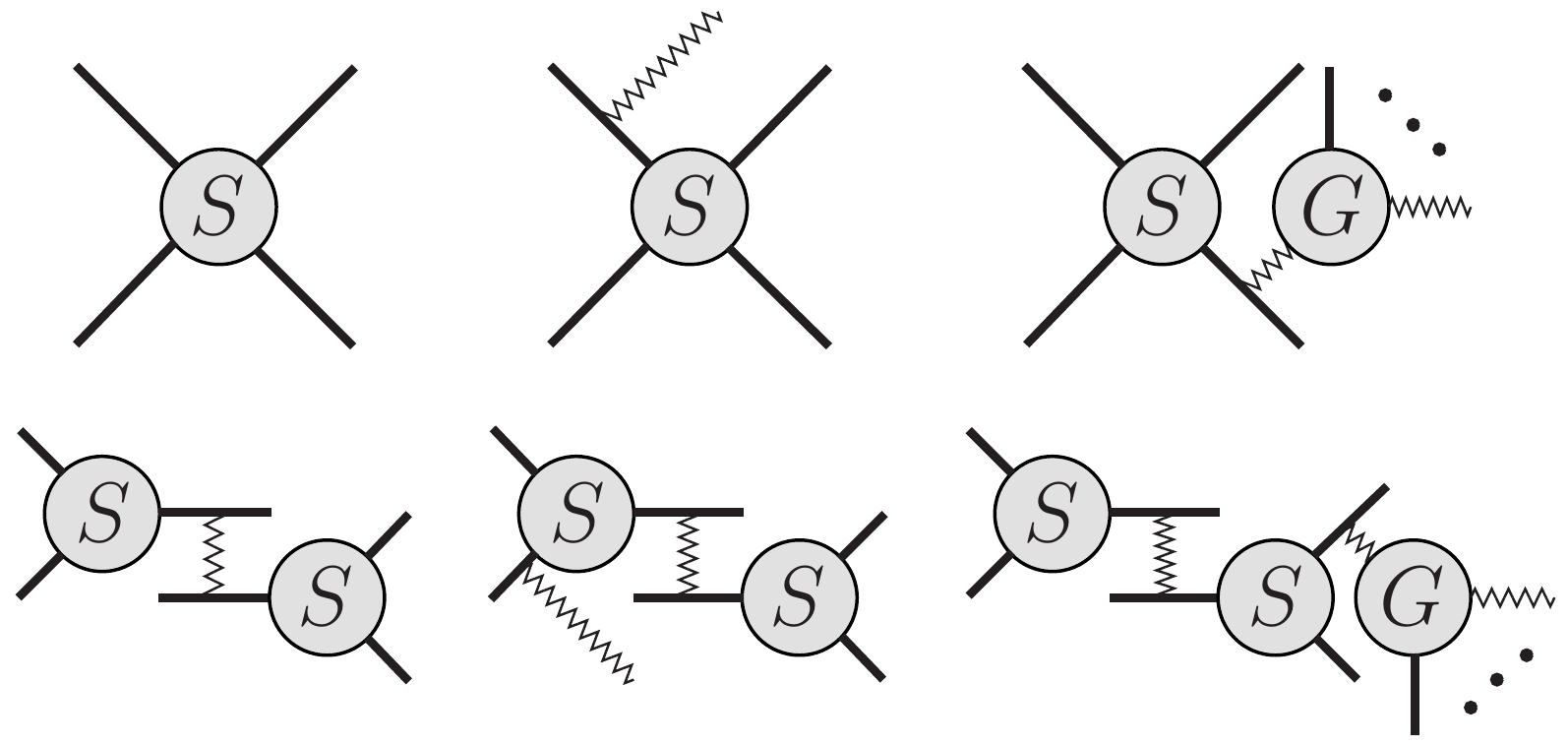}
\caption{Amplitudes of $D=0$.} \labell{fig-4a}
\end{center}
\end{figure}

Therefore, (pseudo) polynomials come from the first and third diagrams, as listed in Figure \ref{fig-4b}.
The first three diagrams in Figure \ref{fig-4b} admit polynomial 1, while the fourth diagram admits
pseudo polynomial $([34]/\<12\>)^{\pm1}$, as the Yukawa interaction permits the two fermions of its vertex
to have the same helicities. The fourth diagram cannot have a gravitational component
attached to it while maintaining the four same helicities, since fermions and gauge bosons must
appear \textit{in pairs of opposite helicities}
when coupling with gravitons. Finally, since four gauge bosons must take the MHV configuration,
$([34]/\<12\>)^{\pm2}$ is excluded.

\begin{figure}
\begin{center}
\includegraphics[width=1\textwidth]{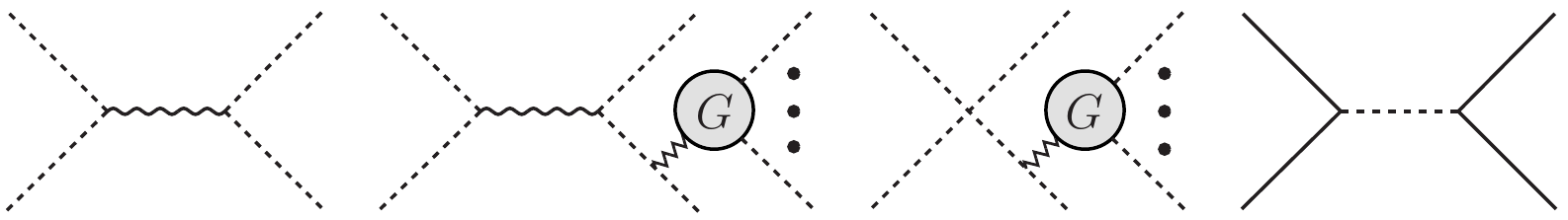}
\caption{(Pseudo) polynomials of $D=0$.} \labell{fig-4b}
\end{center}
\end{figure}

All amplitudes of $D=0$ diagrams other than those in Figure \ref{fig-4b} are solvable.
We find it convenient to attach a gravitational
component to a known diagram to simplify the discussion. This one-line
attachment is equivalent to maximally separating gravitational and Standard Model vertices into two
sub-diagrams, when building the simplest representative diagram.

$\mathbf{D=1}$ \textbf{case}:
Continuing in the same fashion for $D=1$, all possible amplitudes are listed in Figure \ref{fig-5a}.
Here, the 3-point Standard Model vertex can be one of the following four types:
(a) 3-gauge interaction $(\pm1,+1,-1)$; (b) gauge-fermion-fermion interaction $(\pm1,+1/2,-1/2)$;
(c) gauge-scalar-scalar interaction $(\pm1,0,0)$; (d) scalar-fermion-fermion (Yukawa)
interaction $(0,+1/2,-1/2)$.

\begin{figure}
\begin{center}
\includegraphics[width=0.6\textwidth]{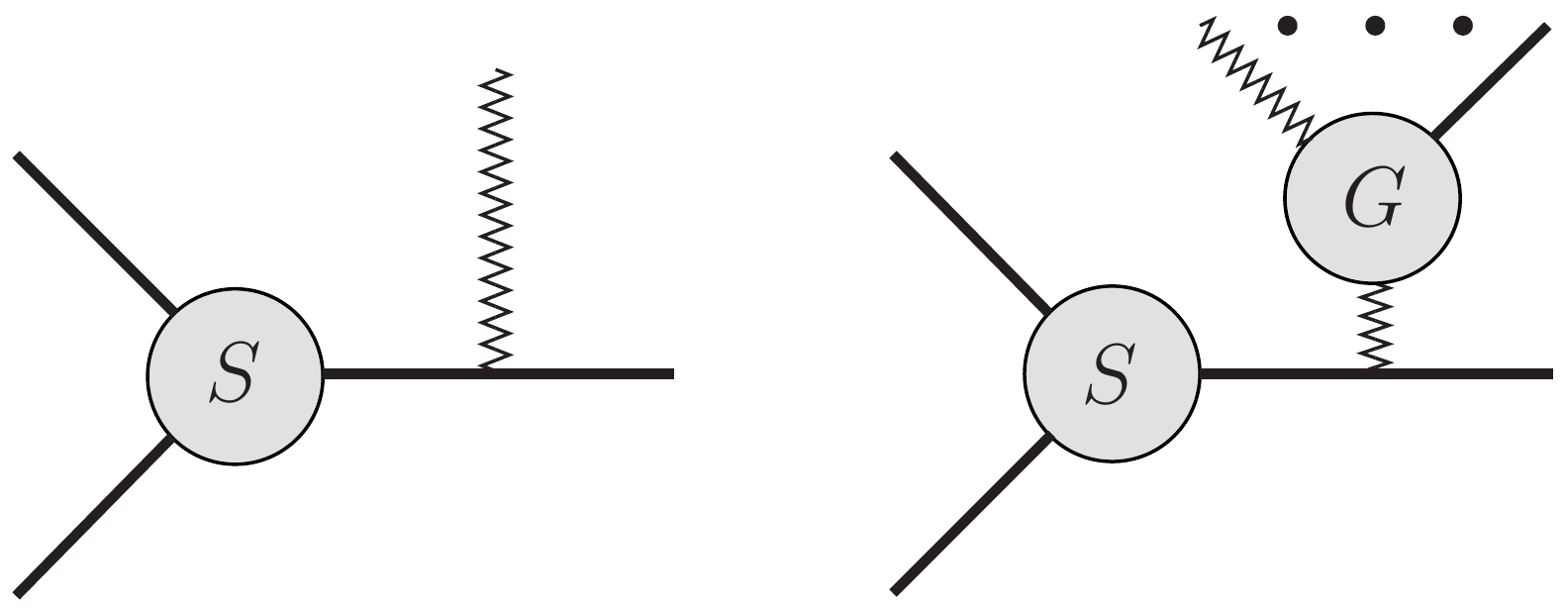}
\caption{Amplitudes of $D=1$.} \labell{fig-5a}
\end{center}
\end{figure}

For the left diagram in Figure \ref{fig-5a},
when vertices of type (a), (b), (c) and (d) are attached by a graviton,
its four helicities are $(\pm1,+1,-1,\pm2)$, $(\pm1,+1/2,-1/2,\pm2)$, $(\pm1,0,0,\pm2)$ and $(0,+1/2,-1/2,\pm2)$
respectively\footnote{Here $\pm1$ and $\pm2$ are independent, they do not necessarily have the same sign.}.
Plugging the data in \eqref{eq-11}, corresponding $D'_{\min}$'s are 3, 3, 3 and 2,
which exclude all four cases. In one words, the left diagram is excluded simply due to the single graviton.

For the right diagram in Figure \ref{fig-5a},
when vertices of type (a), (b) and (c) are attached by a gravitational component, in the most
conservative situation this component only contains external scalars, since higher-spin particles must appear
in pairs of opposite helicities, which will not decrease $D'_{\min}$.
Its helicities are $(\pm1,+1,-1,0,0,\ldots)$, $(\pm1,+1/2,-1/2,0,0,\ldots)$ and $(\pm1,0,0,0,0,\ldots)$
respectively, and $\ldots$ denotes more scalars besides the minimal five. Applying \eqref{eq-11},
corresponding $D'_{\min}$'s are 3, 2 and 1, which exclude first two cases. However,
the third case is also excluded even if its $D'_{\min}$ is allowed. The argument is that no
spinorial product can be formed by only $|1\>^2$ or $|1]^2$, which is known as the Spinor Excess
of inconsistency elimination.
The only polynomial comes from the vertex of type (d), as given in Figure \ref{fig-5b}.
This polynomial is $\<12\>$ or $[12]$.

\begin{figure}
\begin{center}
\includegraphics[width=0.3\textwidth]{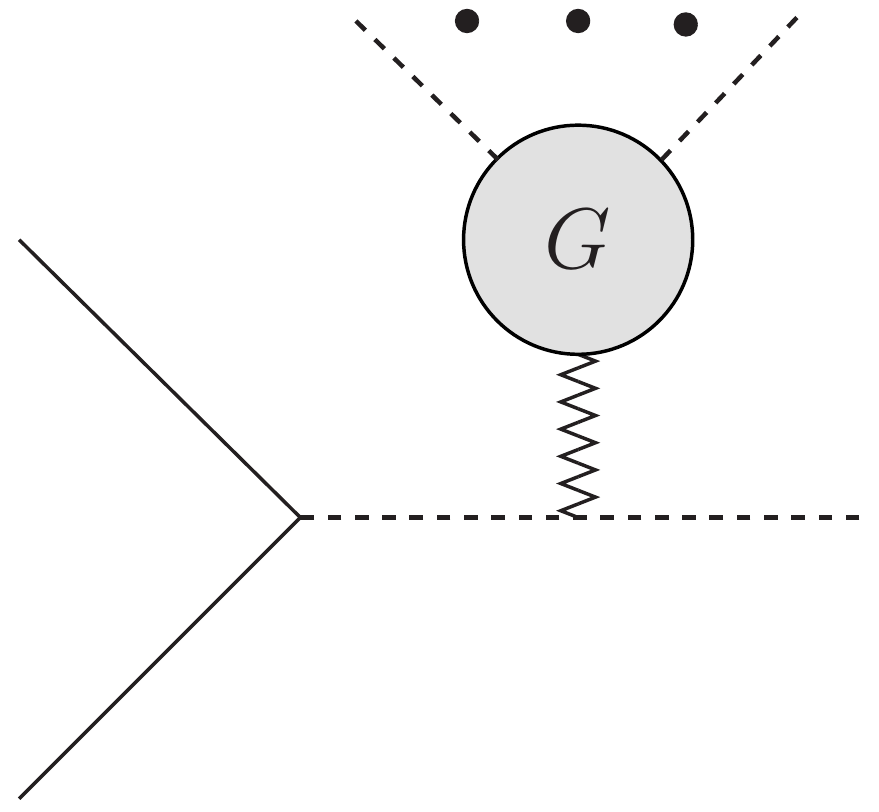}
\caption{Polynomials of $D=1$.} \labell{fig-5b}
\end{center}
\end{figure}

$\mathbf{D=2}$ \textbf{case}:
The last case is $D=2$. The possible amplitude is given in Figure \ref{fig-6a},
and corresponding polynomials are listed in Figure \ref{fig-6b}.
The first one is $P^2_{xy}$, where $x,y$ are two unspecified scalars.
The second one is $\<1x\>[x2]$, with one pair of fermions of opposite helicities.
The third one is $\<12\>[34]$, with two pairs of fermions of opposite helicities.
One may consider a fourth one, with one pair of gauge bosons of opposite helicities, which is
allowed since its $D'_{\min}$ is 2.
But this case is also excluded, as no spinorial product can be formed by only
$|1\>^2|2]^2$ or $|1]^2|2\>^2$.

\begin{figure}
\begin{center}
\includegraphics[width=0.25\textwidth]{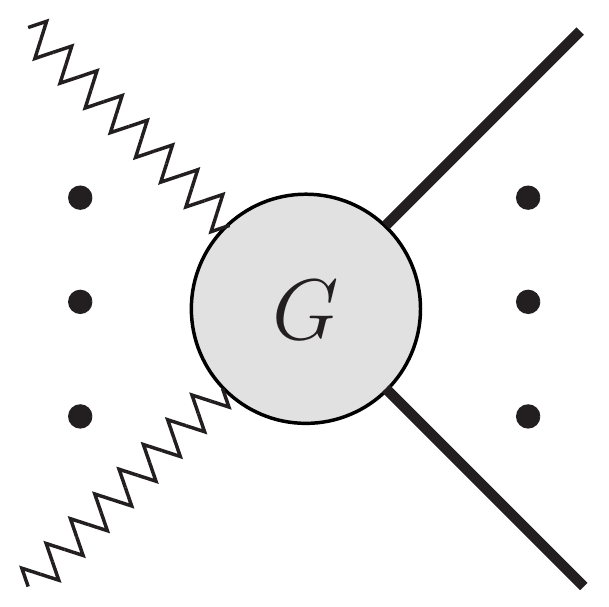}
\caption{Amplitudes of $D=2$.} \labell{fig-6a}
\end{center}
\end{figure}

\begin{figure}
\begin{center}
\includegraphics[width=0.75\textwidth]{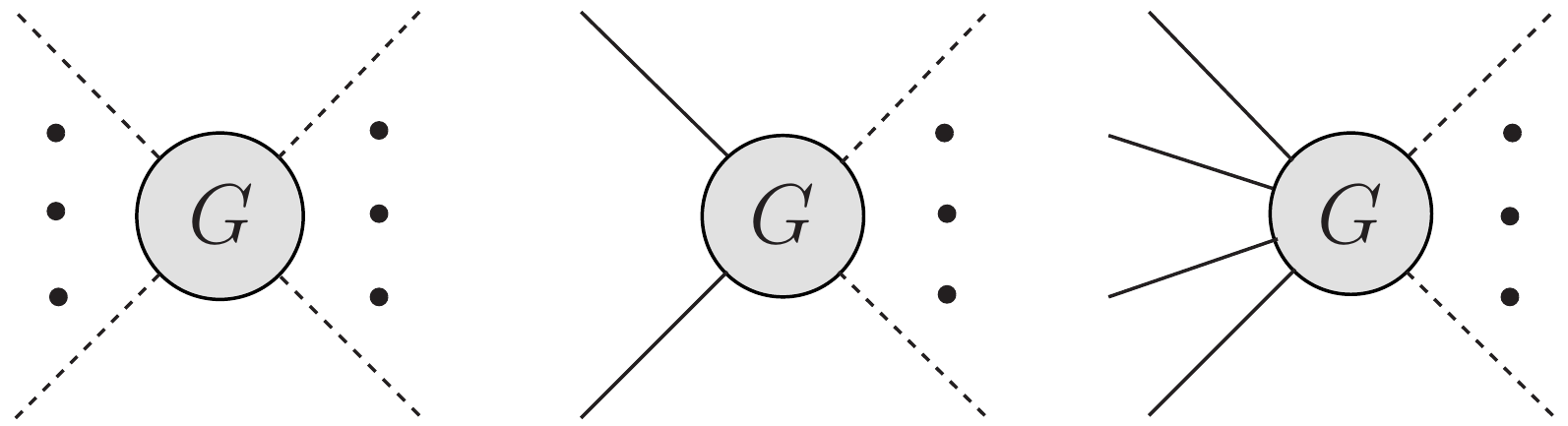}
\caption{Polynomials of $D=2$.} \labell{fig-6b}
\end{center}
\end{figure}

Note that these $D=2$ polynomials can be of either $n=4$, or $n\geq5$ for which all unspecified particles
are scalars. For $n=4$, there are dimensionless pseudo polynomials of the form $([34]/\<12\>)^x$,
which can be an additional factor of the polynomials above.
This factor will lead to a global shift of all four helicities.
But incidentally, there is no extra legitimate pseudo polynomial.

Last but not the least, we need to check the analysis above has covered all possible diagrams,
by using a compact formula
\be
D=2-\sum_{i=1}^S(s_i-2), \labell{eq-12}
\ee
where $D$ is the kinematic mass dimension of a Standard Model `skeleton',
as shown in Figure \ref{fig-7}. This skeleton amplitude
is made of $S$ Standard Model components connected by internal gravitons and $s_i\geq3$
is the number of external legs for each component. When $s_i=2$, there is no Standard Model vertex,
and each component reduces to a Standard Model line,
which should be excluded by the skeleton's definition.

\begin{figure}
\begin{center}
\includegraphics[width=0.4\textwidth]{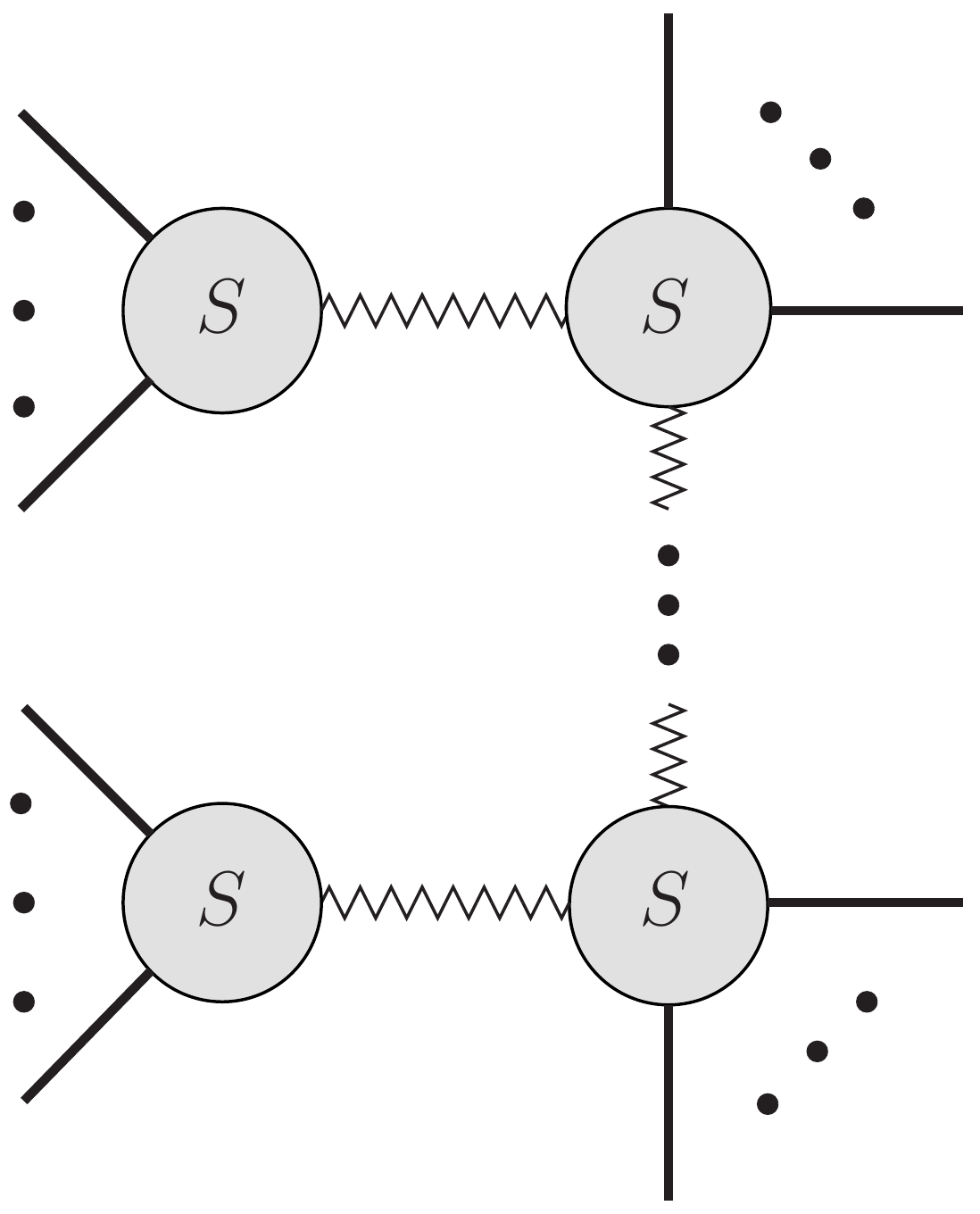}
\caption{Standard Model skeleton. In this case, the unspecified lines can be Standard Model particles only.} \labell{fig-7}
\end{center}
\end{figure}

The proof of this formula is simple. For $S$ connected components, there are $(S-1)$ internal
gravitons. Each internal graviton has two attached points, which bring in two $\kappa$'s.
Hence from \eqref{eq-7}, we have
\be
D=4-\sum_{i=1}^Ss_i-(-2)(S-1),
\ee
which is identical to \eqref{eq-12} after a simple rearrangement.

Having this skeleton, to build a general amplitude, more gravitational components
(either nontrivial components or single gravitons) can be attached to it. By applying the
simplified diagrammatic rules, they can be packed into one single gravitational component.

Now let's consider $D=0$, we have $S=1$, $s_1=4$ and $S=2$, $s_1=s_2=3$.
These two cases correspond to the first and fourth diagrams in Figure \ref{fig-4a}.
When attached by gravitational components, they become the rest four diagrams.
For $D=1$, we only have $S=1$, $s_1=3$. This diagram cannot stand alone, since there is no physical massless
3-particle amplitude. Hence it must be accompanied by gravitational components, which gives the two
diagrams in Figure \ref{fig-5a}.
For $D=2$, the amplitude contains only gravitational vertices, which corresponds to the diagram
in Figure \ref{fig-6a}. Therefore all possible diagrams have been covered.

\section{Discussions}
\labell{discuss}

In this work, we are mainly concerned with the workability of
multi-step BCFW recursion relations. The key techniques of this approach are pole concentration
and inconsistency elimination. Its applicable range has been clarified and we find three
types of objects invulnerable to BCFW deformations: polynomials, pseudo polynomials and
saturated fractions. While the last two objects can be determined by other types of deformations,
how to deal with the first one is probably beyond usual on-shell methods and it may lead to important
generalizations of the present approach. Moreover, when saturated fractions arise, we need to
discuss whether they are legitimate, if not, finding an argument to rule them out is a valuable topic.
Again, we would like to emphasize that this systematic algorithm uses general properties of
field theories only, such as Lorentz invariance, locality and unitarity.
The major information we use are mass dimension and helicities.

Ensuring its workability, we try to further improve the efficiency of multi-step BCFW recursion relations
by taking two sophisticated aspects into account, as listed below:

(a) Knowing the (final or intermediate) boundary term's schematic form (in terms of un-contracted spinors),
it is also very natural to seek for a deformation which renders the boundary term vanish under the large $z$
limit, other than employing inconsistency elimination. In practice, one can consider both ways at each step
to shorten the sequence of pole concentration. More profoundly, inconsistency elimination is only an
argument \textit{afterwards}, as any boundary term that vanishes must be killed by a \textit{good} deformation
with respect to that step.

(b) In practice, it is often evident that there is no need to reach the \textit{final} denominator \eqref{eq-5}.
Merely a particular intermediate form is sufficient to complete the calculation correctly.
This is due to the fact that pole concentration is for \textit{eliminating as many neutral momenta as possible}
in the denominator, and hence increasing the `net spinors' in the numerator while fixing the helicities.

Let's illustrate these two aspects through three simple examples.
The first one is the MHV amplitude $A(1^-,2^+,3^-,4^+,\ldots,n^+)$, where $1^-$ and $3^-$ are non-adjacent.
Assuming all lower-point MHV amplitudes are known, non-vanishing
factorization limits give the common denominator
\be
\frac{1}{\<12\>\<23\>\ldots\<n-1,n\>\<n1\>},
\ee
on the other hand, the kinematic dimension of this amplitude is $(4-n)$, as the gauge coupling constant
is dimensionless. The correct helicities require its numerator to be $|1\>^4|3\>^4$ schematically,
which uniquely fixes the amplitude as
\be
\frac{\<13\>^4}{\<12\>\<23\>\ldots\<n-1,n\>\<n1\>},
\ee
from this example, we see the schematic form is a simple but powerful tool.

The second example is amplitude $A(1^{-1},2^{+1},3^{-1},4^{+1},5^{-2})$ in Einstein-Maxwell theory.
Non-vanishing factorization limits give all physical poles as $[12][32][14][34][15][25][35][45]$.
Since its kinematic dimension is 2, the correct helicities require its schematic form to be
\be
\frac{[1|~[2|^5[3|~[4|^5\times\prod^4|\b\>[\b|}{[12][32][14][34][15][25][35][45]},
\ee
where $|\b\>[\b|$'s are unspecified neutral momenta. At this stage, one can already check that
\be
\<1|5]~~~\<2|5]~~~\<3|5]~~~\<4|5]
\ee
are \textit{good} deformations since each of them induces $z^3$ in the denominator, while the numerator
can at most contain $z^2$ to avoid Spinor Excess, \ie $\prod^4|\b\>$ in the numerator
can at most contain two identical spinors to form non-vanishing spinorial products,
which restricts $\prod^4[\b|$ in the same way. For this case, to find a deformation with
maximal large $z$ suppression is clearly more straightforward than using inconsistency elimination
after a number of constant extractions.

The third example is even more interesting. Consider amplitude $A(1^{-1},2^{+1},3^{-1},4^{+1},5^{-1},6^{+1})$
again in Einstein-Maxwell theory. Factorization limits, mass dimension and helicities together fix its
schematic form to be
\be
\frac{\<1|^2[2|^2\<3|^2[4|^2\<5|^2[6|^2\times\prod^{32}|\b\>[\b|}{P_{12}^2P_{14}^2P_{16}^2
P_{32}^2P_{34}^2P_{36}^2P_{52}^2P_{54}^2P_{56}^2\times P_{132}^2P_{134}^2P_{136}^2
P_{152}^2P_{154}^2P_{156}^2P_{352}^2P_{354}^2P_{356}^2},
\ee
and we will show that two successive deformations, namely
\be
\<3|1]\to\<5|1]
\ee
can already capture the full amplitude. First, after constant extraction $\<3|1]$, similar trick of
pole concentration gives its schematic form
\be
\frac{\<1|^{14}[2|^2[3|^{10}[4|^2\<5|^2[6|^2\times\prod^{22}|\b\>[\b|}
{P_{52}^2P_{54}^2P_{56}^2P_{132}^2P_{134}^2P_{136}^2\<12\>^2\<14\>^2\<16\>^2[32]^2[34]^2[36]^2
\<1|5+2|3]^2\<1|5+4|3]^2\<1|5+6|3]^2},
\ee
and constant extraction $\<5|1]$ turns it into
\be
\frac{\<1|^{20}[2|^2[3|^{10}[4|^2[5|^4[6|^2\times\prod^{18}|\b\>[\b|}
{\<12\>^3\<14\>^3\<16\>^3[32]^2[34]^2[36]^2[52][54][56]\<1|3+2|5]\<1|3+4|5]\<1|3+6|5]
\<1|5+2|3]^2\<1|5+4|3]^2\<1|5+6|3]^2},
\ee
then there is no need to proceed further, because in the numerator Spinor Excess already arises, as
$\prod^{18}|\b\>$ can never saturate $\<1|^{20}$ to form non-vanishing spinorial products.
By this way, two steps can already get the correct answer, while a blind pole concentration in general
requires $4(6-3)=12$ steps. Therefore, it is not always necessary to reach the final denominator,
when eliminating part of neutral momenta in the denominator enforces the numerator
to contain sufficient net spinors for triggering Spinor Excess.

But this is not the end of the story. When proceeding with the calculation
\be
I=P_{\<5|1]}+C_{\<5|1]}P_{\<3|1]}+C_{\<5|1]}C_{\<3|1]},
\ee
while it is just shown that $C_{\<5|1]}C_{\<3|1]}=0$, incidentally we also find $C_{\<5|1]}P_{\<3|1]}=0$.
This means $\<5|1]$ is a \textit{good} deformation and hence one step is enough.
Since particles $1^{-1}$, $3^{-1}$ and $5^{-1}$ are symmetric in the helicity configuration
(there is no color order), $\<3|1]$ is also a good deformation.

In general, there is a \textit{last good deformation} corollary:
After the $n$-th step, when $I=(\textrm{known terms})_n+C_n\cdots C_0$ is reached,
we can further expand it by one more step as
\be
I=P_{n+1}+C_{n+1}(\textrm{known terms})_n+C_{n+1}C_n\cdots C_0,
\ee
assume the $(n+1)$-th step is the last step, for which $C_{n+1}C_n\cdots C_0=0$, and if
\be
C_{n+1}(\textrm{known terms})_n=0,
\ee
then the $(n+1)$-th step is a good deformation. This corollary is powerful in practical calculations,
since unnecessary steps can be saved if we incidentally encounter the condition above.

Back to the mainline, these two aspects (a) and (b) will be demonstrated more systematically,
with more examples in our future work, with a possible joint use of
the last good deformation corollary.
The major goal is to improve the efficiency provided the workability is ensured. The exit of this maze is
now found, and how to shorten the correct route is a complicated yet fascinating problem.
Finally, we would like to highlight the power of simple analysis by mass dimension and helicities, as these
cheap information possibly lie in the core of the future study of efficiency.

\section*{Acknowledgement}

The authors would like to thank Qingjun Jin and Rijun Huang for valuable discussions.
JR is grateful to Qingjun Jin for correcting the errors of the early manuscript.
This work is supported by Qiu-Shi funding and
Chinese NSF funding under contracts No.11031005, No.11135006 and No.11125523.

\appendix
\section{Independent Kinematic Variables}
\labell{app1}

In the calculation of amplitudes, it is common that two visually different expressions in terms of
spinorial products are in fact equivalent. Although one can use a numerical method to check this equivalence,
it is still favorable to find an analytic way.

For an $n$-particle amplitude, let's start with all holomorphic spinorial products as listed below:
\be
\barr{ccccccc}
\<12\> & {} & {} & {} & {} & {} & {} \\
\<13\> & \<23\> & {} & {} & {} & {} & {} \\
\<14\> & \<24\> & \<34\> & {} & {} & {} & {} \\
\<15\> & \<25\> & \<35\> & \<45\> & {} & {} & {} \\
\vdots & \vdots & \vdots & \vdots & \ddots & {} & {} \\
\<1,n-1\> & \<2,n-1\> & \<3,n-1\> & \<4,n-1\> & \ldots & \<n-2,n-1\> & {} \\
\<1n\> & \<2n\> & \<3n\> & \<4n\> & \ldots & \<n-2,n\> & \<n-1,n\>
\earr
\ee
We can pick $|1\>,|2\>$ as two reference spinors, and
for $i,j\neq1,2$ all independent Schouten identities can be solved via
\be
\<ij\>=\frac{1}{\<12\>}\lv\barr{cc}
\<1i\> & \<2i\>\\
\<1j\> & \<2j\>
\earr\rv.
\ee
There are $C^2_n$ $\<ij\>$'s and $C^2_{n-2}$ Schouten identities, so there are
$C^2_n-C^2_{n-2}=2n-3$ independent $\<ij\>$'s.

We can repeat this for anti-holomorphic spinorial products,
and get $(4n-6)$ independent kinematic variables as listed below:
\be
\barr{ccccccccc}
\<12\> & {} & {} & {} & [12] & {} \\
\<13\> & \<23\> & {} & {} & [13] & [23] \\
\<14\> & \<24\> & {} & {} & [14] & [24] \\
\vdots & \vdots & {} & {} & \vdots & \vdots \\
\<1,n-1\> & \<2,n-1\> & {} & {} & [1,n-1] & [2,n-1] \\
\<1n\> & \<2n\> & {} & {} & [1n] & [2n]
\earr
\ee
But the momentum conservation has not been imposed yet.
Adding this constraint, we can solve for $[13]$, $[14]$, $[23]$ and $[24]$ for example, via
\be
\bal
\<1|\sum P|1]=0:~~~\<13\>[13]+\<14\>[14]&=-\Sigma_{11}-\<12\>[12],\\
\<2|\sum P|1]=0:~~~\<23\>[13]+\<24\>[14]&=-\Sigma_{21},\\
\<1|\sum P|2]=0:~~~\<13\>[23]+\<14\>[24]&=-\Sigma_{12},\\
\<2|\sum P|2]=0:~~~\<23\>[23]+\<24\>[24]&=-\Sigma_{22}-\<12\>[12],
\eal
\ee
where $\Sigma_{ij}=\sum^n_{k=5}\<ik\>[jk]$. Now there are $(4n-10)$ independent kinematic variables left.
Any expression in terms of the remaining variables is uniquely fixed.

\section{Example of Pole Concentration}
\labell{app2}

Here, we introduce one systematic sequence to turn all poles of the final boundary term into the
common denominator $\<i_1i_2\>^m[i_3i_4]^{\m}$.
In this example, there are four series of BCFW constant extractions, namely I, II, III and IV:
\be
\bal
\textrm{I}:~~~&\<2|3]~~~&&\<3|4]~~~&&\ldots~~~&&\<n-2|n-1]~~~&&\<n-1|n]\\
\textrm{II}:~~~&\<1|n-1]~~~&&\<2|n-1]~~~&&\ldots~~~&&\<n-3|n-1]~~~&&\<n-2|n-1]\\
\textrm{III}:~~~&\<3|2]~~~&&\<4|3]~~~&&\ldots~~~&&\<n-3|n-4]~~~&&\<n-2|n-3]\\
\textrm{IV}:~~~&\<n-3|1]~~~&&\<n-3|2]~~~&&\ldots~~~&&\<n-3|n-5]~~~&&\<n-3|n-4]\\
\eal
\ee
where III and IV and can be copied from I and II, by reducing $n$ to $(n-2)$
and swapping the holomorphic and anti-holomorphic deformed spinors.
They have $4(n-3)$ steps in total.

To see how this fully works, we assume that all physical poles occur (restrictions such as color order,
must be disregarded). For convenience, let's define sets $H=\<1,\ldots,n\>$ and $A=[1,\ldots,n]$
to denote all holomorphic and anti-holomorphic poles of $n$ particles respectively.
Note that $H$ and $A$ are not fixed, but change at each step of deformation.

Also, the set of all multi-particle poles is denoted by $M$. To fit the analysis of pole concentration,
we will classify all multi-particle poles according to which particle is absent, in a default order.
To be concrete, they are listed as
\be
(p_2+X_{23})^2~~~(P_{23}+X_{24})^2~~~(P_{24}+X_{25})^2~~~\ldots~~~
(P_{2,n-4}+X_{2,n-3})^2~~~(P_{2,n-3}+X_{2,n-2})^2~~~(P_{2,n-2}+X_{2,n-1})^2
\ee
where $P_{ij}=p_i+\ldots+p_j$, $X_{ij}$ is a sum of external momenta without those from $p_i,\ldots,p_j$
and $i,\ldots,j$ is the default order. As the pole momentum includes at least three particles,
$X_{23}$ must at least include two particles, $X_{24}$ must at least include one and
$X_{25}$ can be empty. Similarly, $X_{2,n-3}$ must at most include two particles,
$X_{2,n-2}$ must at most include one and $X_{2,n-1}$ must be empty. Analogous restriction
holds for all $X_{2i}$'s in between. In this list, $p_2$ is the pivot momentum which is always included,
and $p_3,\ldots,p_{n-1}$ becomes the absent momentum one by one.

Now for $\<2|3]$ in series I, the affected two-particle poles are
\be
\bal
\<2\b\>&\not\in H\textrm{ ex }\<23\>\\
[\b3]&\not\in A\textrm{ ex }[23]\\
\eal
\ee
where `ex' means `except', namely all $\<2\b\>$ poles are filtered out except $\<23\>$,
and all $[\b3]$ poles are filtered out except $[23]$. The affected multi-particle pole in $M$ is
\be
(p_2+X_{23})^2\to[2|X_{23}|3\>
\ee
since $(p_2+X_{23})^2$ includes particle 2 but not 3, it is $\<2|3]$ detectable, while all other
multi-particle poles are not, hence they remain unchanged.

Then, for $\<3|4]$ we have
\be
\bal
\<3\b\>&\not\in H\textrm{ ex }\<34\>&&\<23\>\to\<24\>\\
[\b4]&\not\in A&&[\b3]\in A\textrm{ ex }[34]\\
\eal
\ee
where $\<23\>$ turns into $\<24\>$ in $H$ and all $[\b3]$ poles except $[34]$ are revived, which gives
a net effect that all $[\b4]$ poles are filtered out. Now the second multi-particle pole affected is
\be
(P_{23}+X_{24})^2\to[3|p_2+X_{24}|4\>
\ee
again, since $(P_{23}+X_{24})^2$ includes particle 3 but not 4, it is $\<3|4]$ detectable, while remaining
multi-particle poles are unchanged. But don't forget the spurious pole from $\<2|3]$, namely
\be
[2|X_{23}|3\>\to[2|X_{24}|4\>\textrm{ or }[2x_{24}]\<x_{24}4\>
\ee
here when $X_{23}$ includes two particles and one is 4 which is filtered out by $|4\>$, we are left with
one particle $x_{24}$, where $x_{ij}$ denotes one external momentum except that from $p_i,\ldots,p_j$.
A nice fact is that the split poles $[2x_{24}]$ and $\<x_{24}4\>$ are not the ones already
filtered out by $\<2|3]$ and $\<3|4]$, in other words, the next steps will take care of them so there is
no need to look back.

Continuing in this fashion, for $\<4|5]$ we have
\be
\bal
\<4\b\>&\not\in H\textrm{ ex }\<45\>~~\<34\>\to\<35\>~~\<24\>\to\<25\>\\
[\b5]&\not\in A\textrm{ (no ex)}
\eal
\ee
and
\be
\bal
{[2|}X_{24}|4\>&\to[2|X_{25}|5\>\textrm{ or }[2x_{25}]\<x_{25}5\>\\
[3|p_2+X_{24}|4\>&\to[3|p_2+X_{25}|5\>\textrm{ or }[32]\<25\>\\
(P_{24}+X_{25})^2&\to[4|P_{23}+X_{25}|5\>
\eal
\ee
again the split two-particle poles will be taken care of by the next steps.
Note the descendent poles from $(P_{24}+X_{25})^2,\ldots,(P_{2,n-3}+X_{2,n-2})^2,P_{2,n-2}^2$
will no longer produce split poles. After step $\<n-2|n-1]$, all descendent poles from $M$
except split poles are
\be
\bal
&{[2|}X_{2,n-1}|n-1\>&&[3|p_2+X_{2,n-1}|n-1\>&&[4|P_{23}+X_{2,n-1}|n-1\>~~~\ldots\\
&[n-4|P_{2,n-5}+X_{2,n-1}|n-1\>&&[n-3|P_{2,n-4}+X_{2,n-1}|n-1\>&&[n-2|P_{2,n-3}|n-1\>
\eal
\ee
and the last step $\<n-1|n]$ will turn them into (split poles are neglected)
\be
\bal
&{}&&[3|P_{12}|n\>&&[4|P_{13}|n\>\textrm{ or }[4|P_{23}|n\>~~~\ldots\\
&[n-4|P_{1,n-5}|n\>\textrm{ or }[n-4|P_{2,n-5}|n\>
&&[n-3|P_{1,n-4}|n\>\textrm{ or }[n-3|P_{2,n-4}|n\>&&[n-2|P_{2,n-3}|n\>
\eal
\ee
For two-particle poles, after series I we have
\be
H=\{\<\b n\>\},~A=[1,\ldots,n-1],
\ee
so all poles in $H$ include $|n\>$ and any pole in $A$ does not include $|n]$.
Of course, all these poles can have orders larger than one.

After series II, it is easy to check that
\be
H=\{\<n-1,n\>\},~A=[1,\ldots,n-2].
\ee
For spurious poles of the form $[\b|\b|\b\>$ from $M$, series II also nicely turns them into
the poles in $H$ and $A$. To verify this, let's single out the following series in II
(note that $\<1|n-1]$ and $\<n-2|n-1]$ are set aside temporarily):
\be
\<2|n-1]~~~\ldots~~~\<n-3|n-1]
\ee
and it will be enough to take care of spurious poles $[\b|\b|\b\>$. Explicitly, we have
\be
\bal
\<2|n-1]:&~~~[3|P_{12}|n\>&&\to[32]\<n-1,n\>\\
\<3|n-1]:&~~~[4|P_{13}|n\>\textrm{ or }[4|P_{23}|n\>&&\to[43]\<n-1,n\>\\
\vdots~~~~~~&\\
\<n-4|n-1]:&~~~[n-3|P_{1,n-4}|n\>\textrm{ or }[n-3|P_{2,n-4}|n\>&&\to[n-3,n-4]\<n-1,n\>\\
\<n-3|n-1]:&~~~[n-2|P_{2,n-3}|n\>&&\to[n-2,n-3]\<n-1,n\>
\eal
\ee
therefore these spurious poles finally become parts of $H$ and $A$ after series I and II,
we are only left with $\{\<n-1,n\>\}$ and $[1,\ldots,n-2]$. To concentrate them completely,
series III and IV are needed, which are copied from I and II by replacing $n$ by $(n-2)$
and swapping $|\b\>$ and $|\b]$ for all deformations.

After series III and IV, a trivial imitation gives
\be
H=\{\<n-1,n\>\},~A=\{[n-3,n-2]\}.
\ee
Note that $H$ is in fact completely inert to series III and IV as only $A$ is manipulated.
Therefore we manage to turn all poles, regardless of whether they are of two or multiple particles,
physical or spurious, into a common denominator
\be
\frac{1}{\<i_1i_2\>^m[i_3i_4]^{\m}},
\ee
where $i_1,i_2,i_3,i_4$ are four different arbitrary particles after a trivial relabeling.

For $n=4$, merely series I and II can turn all poles into such a form
(in fact series III and IV do not exist). Consider the denominator $P_{23}^2P_{24}^2P_{34}^2$,
as there are many equivalent choices related by momentum conservation, after
\be
\<2|3]~~~\<3|4]~~~\<1|3]~~~\<2|3]~~~
\ee
it becomes $\<34\>^3[12]^3$. Again a relabeling gives the general form $\<i_1i_2\>^3[i_3i_4]^3$.

In the sequence above, series I and III are skew descendant while series II and IV are straight descendent.
In general, not all physical poles can be detected by merely independent and straight descendent deformations.
To calculate an $n$-particle amplitude with only these two types, one can assign $a$ labels for $\<i|$
and $(n-a)$ labels for $|j]$ in a sequence $\<i|j]$. Then two-particle poles $[i_1i_2]$
with $i_1,i_2\in I_a$ and $\<j_1j_2\>$ with $j_1,j_2\in I_{n-a}$ cannot be detected, where
$I_a$ and $I_{n-a}$ denote the sets of $\<i|$ and $|j]$ respectively.

\section{(Pseudo) Polynomials and Saturated Fractions}
\labell{app3}

This part presents the classification of all three types of objects that cannot be determined by
multi-step BCFW recursion relations. They include (pseudo) polynomials and saturated fractions.

\subsection{Polynomials and pseudo polynomials}

Now, we list all (pseudo) polynomials that satisfy certain helicities, up to $D=2$.
This list can be similarly extended for $D\geq3$.

First consider $n\geq5$, as there is a tricky issue of $n=4$. For dimension
(or polynomial degree) $D=0$, there is only one choice of polynomial, namely 1.
The helicity configuration is simply
\be
(0,0,0,0,0,\ldots)
\ee
where $\ldots$ denotes more scalars besides the minimal five.

For dimension $D=1$, there are two choices: $\<\b\b\>$ or $[\b\b]$ and the helicity configuration is
\be
\(\pm\frac{1}{2},\pm\frac{1}{2},0,0,0,\ldots\)
\ee
For dimension $D=2$, there are three choices: $\<\b\b\>\<\b\b\>$,
$\<\b\b\>[\b\b]$ or $[\b\b][\b\b]$.
One also needs to separate the cases for which one or two pair(s) of particle labels
in the spinorial products are identical. Hence the helicity configuration can be
\be
\bal
&\(\pm\frac{1}{2},\pm\frac{1}{2},\pm\frac{1}{2},\pm\frac{1}{2},0,\ldots\)~~~
\(\pm\frac{1}{2},\pm\frac{1}{2},\pm1,0,0,\ldots\)~&&(\pm1,\pm1,0,0,0,\ldots)\\
&\(-\frac{1}{2},-\frac{1}{2},+\frac{1}{2},+\frac{1}{2},0,\ldots\)~~~
\(-\frac{1}{2},+\frac{1}{2},0,0,0,\ldots\)~&&(0,0,0,0,0,\ldots)
\eal
\ee
When $n=4$, there exist fraction-like pseudo polynomials.
Explicitly, from the following identity
\be
\frac{[34]}{\<12\>}=-\frac{[32]}{\<14\>}=\frac{[42]}{\<13\>}
=-\frac{[31]}{\<42\>}=\frac{[12]}{\<34\>}=-\frac{[14]}{\<32\>},
\ee
one can check that, there is no effective BCFW deformation that can detect poles above, as the numerator
vanishes simultaneously when the denominator reaches to zero, due to momentum conservation.
Moreover, pseudo polynomials are in fact inert to BCFW deformations, for example under $\<1|3]$,
\be
\frac{[34]}{\<12\>}\to\frac{[34]+z[14]}{\<12\>-z\<32\>}=-\frac{[14]}{\<32\>}\times
\frac{[34]/[14]+z}{-\<12\>/\<32\>+z}=-\frac{[14]}{\<32\>}=\frac{[34]}{\<12\>},
\ee
again due to momentum conservation.
In general, a dimensionless pseudo polynomial takes the form
\be
\(\frac{[34]}{\<12\>}\)^x,
\ee
where $x=\pm1,\pm2,\pm3,\pm4$, as the spin is restricted within 2.

Since the dimensionless pseudo polynomial behaves like a numerical factor, polynomials of $n=4$
that share the same forms as those of $n\geq5$
can be multiplied by these factors to generate all possible pseudo polynomials.
Hence, for $n=4$ we have the following helicity configurations:
\be
\bal
D=0:~~~&(y,y,y,y)~&&-2\leq y\leq2\\
D=1:~~~&\(\frac{1}{2}+y,\frac{1}{2}+y,y,y\)~&&-2\leq y\leq\frac{3}{2}\\
D=2:~~~&(y,y,y,y)~&&-2\leq y\leq2\\
&(1+y,1+y,y,y)~&&-2\leq y\leq1\\
&\(-\frac{1}{2}+y,\frac{1}{2}+y,y,y\)~&&-\frac{3}{2}\leq y\leq\frac{3}{2} \labell{eq-3}
\eal
\ee
where $y=0,x/2$. Since (pseudo) polynomials of $n=4$
are dimensionally equal, up to a global shift of all four helicities,
for simplicity one can first select a set of representatives to analyze.
As one example, the representative helicity configuration can be
\be
\bal
D=0:~~~&(0,0,0,0)\\
D=1:~~~&\(\pm\frac{1}{2},\pm\frac{1}{2},0,0\)\\
D=2:~~~&(0,0,0,0)\\
&(\pm1,\pm1,0,0)\\
&\(-\frac{1}{2},\frac{1}{2},0,0\)
\eal
\ee
then $y$ must be shifted around to cover all possible cases in \eqref{eq-3}.

\subsection{Saturated fractions}

The SF is an irreducible fraction, but impossible to be killed
by BCFW constant extractions. A minimally $z$-power inducing constant extraction
will only change its particle labels, but not its form. For example,
\be
\frac{[34][56]}{\<12\>}
\ee
is a pure SF, and constant extraction $\<1|3]$ turns it into
\be
C_{\<1|3]}\frac{[34][56]}{\<12\>}=-\frac{[14][56]}{\<32\>}.
\ee
For a pure SF, all effective BCFW deformations are minimally $z$-power inducing. Next, consider
\be
\frac{[34]^2[56]}{\<12\>}=[34]\times\frac{[34][56]}{\<12\>},
\ee
which is a mixed SF, namely a pure SF times a polynomial.
When we use $\<1|3]$ again, it is not minimally $z$-power inducing, \ie
\be
\frac{[34]^2[56]}{\<12\>}\to\frac{([34]+z[14])^2[56]}{\<12\>-z\<32\>}\sim O(z),
\ee
for this mixed SF, one minimally $z$-power inducing deformation is $\<1|5]$, for which the constant extraction
will give only one term with relabeling. But $\<1|3]$ gives more than one term, as
\be
C_{\<1|3]}\frac{[34]^2[56]}{\<12\>}=-2[34]\times\frac{[14][56]}{\<32\>}-\frac{[14]^2\<12\>[56]}{\<32\>^2},
\ee
note that the first term above is also a mixed SF again with relabeling.
In general, under a non-minimally $z$-power inducing constant extraction, a mixed SF will transform into
more than one term. These terms include a special one, which is related to the original SF by relabeling.

To investigate SF's more systematically,
we analyze a special category: When $0\leq h_1,h_2<m/2$ and
$0\leq h_3,h_4$, the dependence on $m,\m$ dissolves in \eqref{eq-6}, then
\be
D'=\sum_{i=1,2}(-h_i+\bt_i)+\sum_{i=3,4}(h_i+\ap_i)
+\sum_{h<0}(-h_i+\bt_i)+\sum_{h\geq0}(h_i+\ap_i), \labell{eq-9}
\ee
and the corresponding final boundary term schematically reads
\be
\frac{1}{\<12\>^m[34]^{\m}}\prod_{i=1,2}\<i|^{-2h_i+m}p_i^{\bt_i}\prod_{i=3,4}[i|^{2h_i+\m}p_i^{\ap_i}
\prod_{h<0}\<i|^{-2h_i}p_i^{\bt_i}\prod_{h\geq0}[i|^{2h_i}p_i^{\ap_i}.
\ee
\textbf{1st type:}
Let's consider the first type of amplitudes\footnote{All amplitudes in this appendix are not necessarily
physical, as the investigation is purely of mathematical interest.},
of helicity configuration $(h,h,h,h,h,\ldots)$ where $0\leq h\leq2$.
It is easy to see that SF's of $n=4$ are identical to pseudo polynomials,
so nontrivial SF's only exist for $n\geq5$. At this point, \eqref{eq-9} becomes
\be
D'=(n-4)h+\bt_1+\bt_2+\sum_{i=3}^n\ap_i, \labell{eq-14}
\ee
and the final boundary term schematically reads
\be
\frac{1}{\<12\>^{2h}}\<1|^{\bt_1}\<2|^{\bt_2}\<3|^{\ap_3}\<4|^{\ap_4}\ldots\<n|^{\ap_n}
\times[1|^{\bt_1}[2|^{\bt_2}[3|^{2h+\ap_3}[4|^{2h+\ap_4}\ldots[n|^{2h+\ap_n}, \labell{eq-15}
\ee
where we have maximally reduced $\<1|,\<2|$ in the fraction, hence in the numerator
\be
(\textrm{num. of }|\b\>)=\bt_1+\bt_2+\sum_{i=3}^n\ap_i,~
(\textrm{num. of }|\b])=(n-2)2h+\bt_1+\bt_2+\sum_{i=3}^n\ap_i.
\ee
For this type, FD arises when $n=\textrm{odd}$ and $h=1/2,3/2$ with $D\geq D'_{\min}$,
PM arises when $D'-nh=\textrm{odd}$ with $D\geq D'_{\min}$ and $2nh=\textrm{even}$.
From \eqref{eq-14} and \eqref{eq-15}, it's easy to check that:\\
$~~~~$When $D=0$, there is no consistent choice. We do not consider (pseudo) polynomials here.\\
$~~~~$When $D=1$, $n=5$, $h=1$ is SF, $h=1/2$ is FD, $h=0$ is PM.\\
$~~~~$When $D=1$, $n=6$, $h=1/2$ is SF, $h=0$ is PM.\\
$~~~~$When $D=2$, $n=5$, $h=1$ is PM, $h=1/2$ is FD.\\
$~~~~$When $D=2$, $n=6$, $h=1$ is SF, $h=1/2$ is PM.\\
$~~~~$When $D=2$, $n=7$, $h=1/2$ is FD.\\
$~~~~$When $D=2$, $n=8$, $h=1/2$ is SF.\\
For $D=0,1,2$ all other cases are inconsistent since $D<D'_{\min}$. SF's above are
(the form of an SF is not necessarily unique, so we only present one example for each case)
\be
\bal
&\textrm{SF}(D=1,n=5,h=1)=\frac{[34][35][45]}{\<12\>^2},~
&&\textrm{SF}\(D=1,n=6,h=\frac{1}{2}\)=\frac{[34][56]}{\<12\>},\\
&\textrm{SF}(D=2,n=6,h=1)=\(\frac{[34][56]}{\<12\>}\)^2,~
&&\textrm{SF}\(D=2,n=8,h=\frac{1}{2}\)=\frac{[34][56][78]}{\<12\>}.
\eal
\ee
\textbf{2nd type:}
Similarly, the second type of amplitudes is $(h,h,h+1/2,h,h,\ldots)$ where $0\leq h\leq3/2$, then
\be
D'=(n-4)h+\frac{1}{2}+\bt_1+\bt_2+\sum_{i=3}^n\ap_i.
\ee
For this type, we find that:\\
$~~~~$When $D=0$, there is no consistent choice.\\
$~~~~$When $D=1$, $n=5$, $h=1/2$ is SF, $h=0$ is FD.\\
$~~~~$When $D=1$, $n=6,7,8,\ldots,$ $h=0$ is FD.\\
$~~~~$When $D=2$, $n=5$, $h=3/2$ is SF, $h=1$ is FD, $h=1/2$ is PM, $h=0$ is FD.\\
$~~~~$When $D=2$, $n=6$, $h=3/2,1,1/2,0$ is FD.\\
$~~~~$When $D=2$, $n=7$, $h=1/2$ is SF, $h=0$ is FD.\\
$~~~~$When $D=2$, $n=8,9,10,\ldots,$ $h=0$ is FD.\\
SF's above are
\be
\bal
&\textrm{SF}\(D=1,n=5,h=\frac{1}{2}\)=\frac{[34][35]}{\<12\>},~\\
&\textrm{SF}\(D=2,n=5,h=\frac{3}{2}\)=\frac{[34]^2[35]^2[45]}{\<12\>^3},~
\textrm{SF}\(D=2,n=7,h=\frac{1}{2}\)=\frac{[34][35][67]}{\<12\>}.
\eal
\ee
\textbf{3rd type:}
The third type is $(h,h,h+1/2,h+1/2,h,\ldots)$ where $0\leq h\leq3/2$, then
\be
D'=(n-4)h+1+\bt_1+\bt_2+\sum_{i=3}^n\ap_i.
\ee
For this type, we find that:\\
$~~~~$When $D=0,1$, there is no consistent choice.\\
$~~~~$When $D=2$, $n=5$, $h=1$ is SF, $h=1/2$ is FD, $h=0$ is PM.\\
$~~~~$When $D=2$, $n=6$, $h=1/2$ is SF, $h=0$ is PM.\\
$~~~~$When $D=2$, $n=7,8,9,\ldots,$ $h=0$ is PM.\\
SF's above are
\be
\textrm{SF}\(D=2,n=5,h=1\)=\frac{[34]^2[35][45]}{\<12\>^2},~
\textrm{SF}\(D=2,n=6,h=\frac{1}{2}\)=\frac{[34]^2[56]}{\<12\>}.
\ee
\textbf{4th type:}
The fourth type is $(h,h,h+1,h,h,\ldots)$ where $0\leq h\leq1$, then
\be
D'=(n-4)h+1+\bt_1+\bt_2+\sum_{i=3}^n\ap_i.
\ee
For this type, we find that:\\
$~~~~$When $D=0$, there is no consistent choice.\\
$~~~~$When $D=1$, $n=5,6,7,\ldots,$ $h=0$ is SE.\\
$~~~~$When $D=2$, $n=5$, $h=1$ is SF, $h=1/2$ is FD, $h=0$ is PM.\\
$~~~~$When $D=2$, $n=6$, $h=1/2$ is SF, $h=0$ is PM.\\
$~~~~$When $D=2$, $n=7,8,9,\ldots,$ $h=0$ is PM.\\
SF's above are
\be
\textrm{SF}\(D=2,n=5,h=1\)=\frac{[34]^2[35]^2}{\<12\>^2},~
\textrm{SF}\(D=2,n=6,h=\frac{1}{2}\)=\frac{[34][35][36]}{\<12\>}.
\ee
We will not continue to explore a fifth type, as these toy examples have provided enough intuition
for the generation of SF's. Naturally SF's are highly un-physical, but their existence in some artificial
(effective) theories cannot be excluded. Sometimes even if an SF is admitted, we can use other
arguments to exclude it, such as its denominator contains a spurious pole. Consider
an SF whose denominator is $\<12\>^2$, or $\<12\>$ but the amplitude has vanishing factorization limit
under $\<12\>\to0$, then one needs to delicately remove all dependence on the spurious pole
$\<12\>^2$, or $\<12\>$, in the known terms of \eqref{eq-1} if it appears.


\end{document}